\documentclass[aps,preprint,nofootinbib,groupedaddress,pra,showpacs]{revtex4}

\usepackage{graphics}
\usepackage{graphicx}
\usepackage{amssymb}
\usepackage{amsmath}
\usepackage{amsfonts}

\begin{document}

\title{Study of Simulation Method of Time Evolution of Atomic and Molecular Systems by Quantum Electrodynamics}

\author{Kazuhide Ichikawa}
\author{Masahiro Fukuda}
\author{Akitomo Tachibana} 
\email{akitomo@scl.kyoto-u.ac.jp}
\affiliation{Department of Micro Engineering, Kyoto University, Kyoto 615-8540, Japan}

\date{\today}

\begin{abstract}
We discuss a method to follow step-by-step time evolution of atomic and molecular systems based on 
QED (Quantum Electrodynamics). Our strategy includes expanding the electron field operator by localized wavepackets
to define creation and annihilation operators
and following the time evolution using the equations of motion of the field operator in the Heisenberg picture.
We first derive a time evolution equation for the excitation operator, the product of two creation or annihilation operators,
which is necessary for constructing operators of physical quantities such as the electronic charge density operator. 
We then describe our approximation methods to obtain time differential equations of the electronic density matrix,
which is defined as the expectation value of the excitation operator.
By solving the equations numerically,
we show ``electron-positron oscillations", the fluctuations originated from virtual electron-positron pair creations and annihilations, appear
in the charge density of a hydrogen atom and molecule. 
We also show that the period of the electron-positron oscillations becomes shorter by including the self-energy process, in which the electron emits a photon and then absorbs it again,
and it can be interpreted as the increase in the electron mass due to the self-energy. 
\end{abstract}


% 03.70.+k	Theory of quantized fields (see also 11.10.-z Field theory)
%12.20.-m	Quantum electrodynamics
%31.30.J-	   Relativistic and quantum electrodynamic (QED) effects in atoms, molecules, and ions

%03.65.Pm	Relativistic wave equations
% 11.10.-z	Field theory (for gauge field theories, see 11.15.-q)
% 11.15.Tk	Other nonperturbative techniques

\maketitle

%%%%%%%%%%%%%%%%%%%%%%%%%%%%%%%%%%%%%%%
%
\section{Introduction}  \label{sec:introduction}
%
%%%%%%%%%%%%%%%%%%%%%%%%%%%%%%%%%%%%%%%
The elementary processes of almost every phenomenon in condensed matter physics and chemistry
 can be regarded as 
time evolution of a system which consists of interacting charged particles and photons.
The physical theories which describe such a system are electromagnetism and quantum mechanics,
and their unified theory has already been constructed before the middle of last century as the 
quantum electrodynamics (QED), taking the form of quantum field theory. 
The QED is the most stringently tested theory of physics and its precision is confirmed by 
many types of experiments of elementary particle physics. 
It can be considered as the most successful fundamental physical theory we have. 

However, there has been only limited use of QED in the fields such as condensed matter physics
 and quantum chemistry. 
There are a lot of works to simulate the time evolution of the quantum systems involving light and matter 
by using the time-dependent Schr\"{o}dinger, Dirac, or DFT (density functional theory) equation with
classical electromagnetic fields ({\it i.e.} semi-classical approximation) \cite{Bauer2006, Iwasa2009, Mocken2008,Joachain},
but both matter and light are not treated as quantized fields.
In some simulations, the quantum photon field is used but the matter part and its interaction with the photon
are much more simplified than QED \cite{Loudon,Meystre,Haroche},
or the quantum Dirac field is used for electrons but the interaction is not given by the photon field \cite{Krekora2005,Krekora2006}.
In atomic physics and quantum chemistry, QED is only regarded as a small correction to the Coulomb potential 
which appears in the time-independent Dirac equation \cite{Dyall,Reiher,Lindgren}, 
and not considered in dynamical situations. 

It is true that the approximations above are valid for a broad range of systems so far, 
but the atomic, molecular, and optical physics experiments are in rapid progress recently.
It is now possible to measure, fabricate, and control structures on the atomic and molecular scale
with the advances in nanotechnology.
A single photon and a single electron spin can be measured and manipulated by the current technology of photonics and spintronics \cite{Kimble1998,Claudon2010,Buller2010,Aharonovich2011}.
Moreover, developments in the laser science have made it possible to observe a phenomenon 
at the time scale of femtosecond to attosecond time scales \cite{Drescher2002,Krausz2009}.
Considering such progress in the experiments 
of smaller space-time scales and the most fundamental particle properties, 
it is important for the theoretical side to develop a simulation method based on as an elementary 
theory as possible. 
This is the reason why we try to formulate a time evolution simulation method for
atomic and molecular systems, which consist of electrons, atomic nuclei and photons,
closely following QED in the form of quantum field theory. 

To achieve our goal to develop a method to simulate time evolution of atomic and molecular systems by QED,
we have to overcome some issues which do not appear in the ordinary QED.
When we say the ordinary QED, we mean that it is a relativistic (Lorentz invariant) quantum field theory and
the calculation method of the scattering amplitude is performed by the covariant perturbation theory \cite{Peskin,WeinbergI}.
It implies that the only transition between infinite past (``in-state") and infinite future (``out-state") is concerned 
and the quantum fields in those states (``asymptotic states") can be treated like non-interacting fields.
This in turn makes it possible to apply the perturbative approach by taking the asymptotic states
(where fields are described by non-interacting Hamiltonian) as the unperturbed states 
and interaction as the perturbation. 
Also, the perturbative method is established in a quite systematic way, owing to the Lorentz invariance of the theory.
Although such a method yields physical quantities which can be precisely compared with 
experiments of particle physics in particular, it is not sufficient for our interests.
For simulating time evolution of atomic and molecular systems,
there are mainly three issues we have to face which are not simultaneously concerned in the ordinary QED:
(i) there are atomic nuclei, which are non-relativistic and not elementary particles,
(ii) the matter particles in those systems are in bound states,
(iii) we would like to follow the finite time evolution of the systems step by step.

In fact, there have already been methods which can partly treat these three points within the framework of
quantum field theory. 
First, as for (i), there are effective field theories in which nucleons are treated as quantized fields \cite{Weinberg1990,WeinbergII}, 
but the computation is within the scattering theory, so they are not suited to our needs of (ii) and (iii).
In principle, since the nucleons consist of quarks and gluons which are described by quantum chromodynamics (QCD),
they can be treated in a quantum field theoretic way,
and the method of lattice field theory has been well developed to perform non-perturbative calculation of QCD, 
known as lattice QCD calculation. 
However, it takes too much computational time even to describe a nucleon as a bound state of quarks, and adding 
QED to study atoms is much more unpractical. 
Other shortcomings include that the current lattice field theory is developed only to describes an equilibrium state 
and time evolution cannot be treated. 
Also, in contrast to lattice QCD, lattice QED has a problem that, 
being $U(1)$ gauge theory which does not exhibit asymptotic freedom, 
the continuum limit of lattice spacing cannot be taken.
Second, as for (ii), a well-known technique is the Bethe-Salpeter equation \cite{Salpeter1951}.
In Ref.~\cite{Grotch2002}, the Bethe-Salpeter equation and various other techniques to describe bound states in QED are reviewed, 
but those incorporating (i) or (iii) are not found. 
Finally, as for (iii), some formalisms are known to describe a non-equilibrium state by quantum field theory,
such as closed time path (CTP) formalism \cite{Rammer, Calzetta} and thermo field dynamics \cite{Umezawa}.
In particular, CTP formalism is applied to gauge theory including QED and QCD, but currently not for 
bound state problems. 
This is because the CTP formalism uses systematic perturbative expansion based on the interaction picture,
but we cannot divide the QED hamiltonian into unperturbed and interaction parts when
we consider bound states. 

Therefore, to simulate time evolution of atomic and molecular systems based on QED, we cannot just use
preexisting formalisms as mentioned above.
We briefly explain our proposals \cite{Tachibana2003,Ichikawa2013} to cope with these issues in the following.
First of all, as for (i), we add the atomic nuclei degree of freedom as Schr\"{o}dinger fields, and 
the interaction with the photon field is determined from $U(1)$ gauge symmetry as usual \cite{Tachibana2003}.
Although this deprives the theory of the Lorentz invariance, since we try to follow the finite time evolution of the system,
this would not be a crucial problem. 
Next, as for (ii), in order to describe the bound state, we expand the matter field operators by localized wavepackets,
not by the usual plane waves, and define the creation and annihilation operators \cite{Ichikawa2013}. 
In the case of the Dirac field operator for electrons, for example, we may adopt the stationary solutions of 
the Dirac equation under the existence of external electrostatic field, as the expansion functions.
This is similar to the Furry picture \cite{Furry1951,LLRQT}, but, as described below, 
we do not assume the time-dependence of the operators which is determined by the energy eigenvalues. 
Finally, as for (iii), we follow the time evolution using the equations of motion of the field operator 
in the Heisenberg picture \cite{Ichikawa2013}.
As is mentioned earlier, since we do not have well-defined division between unperturbed and interaction parts
of the QED hamiltonian for the bound state, we cannot work in the interaction picture. 

The setups described above determine the evolution equations of field operators, and 
those of creation and annihilation operators, 
but we have to make further approximations and assumptions to obtain the evolution equations for
the expectation value of physical quantities. 
In Ref.~\cite{Ichikawa2013}, we have studied the time evolution of one of the most basic physical 
quantity operators, the electronic charge density operator, and have discussed approximation methods to obtain
the time evolution of its expectation value. 
The charge density operator is expressed by the product of two creation or annihilation operators,
which is called an excitation operator, but since the time derivative of creation and annihilation operators contains
more than one of these operators, the time differential equation of the excitation operator is not closed.
In other words, differentiating the excitation operator yields operators which cannot be expressed by the excitation operator. 
Such a problem is generic for interacting quantum field theories \cite{Rammer}, and not special to our approach. 
In Ref.~\cite{Ichikawa2013}, we have introduced several approximations to obtain the time evolution equation 
for the expectation value of the excitation operator, which is called a density matrix, and numerically solved
the time evolution of the density matrix. 
The time evolution of the expectation value of the charge density operator has been obtained by
multiplying the density matrix by the expansion functions of the field operator.
Then, we have found that the time evolution of the charge density of a hydrogen atom 
exhibits very rapid oscillations of the period $\approx 1.7 \times 10^{-4}$\,a.u. ($4.1\times 10^{-21}$\,s), 
which corresponds to the inverse of twice the electron mass.
This is interpreted as the fluctuations originated from the virtual electron-positron pair creations and annihilations,
showing the effect of QED, and we have designated the phenomenon as ``electron-positron oscillations". 

In the present paper, we improve one of the approximations employed in Ref.~\cite{Ichikawa2013} with respect to 
the terms which include the photon creation and annihilation operators.
In our formalism, the time derivative of the electron excitation operator has the terms which consist of 
 two creation or annihilation operators sandwiching a photon creation or annihilation 
operator (we call this type of operator by ``$\hat{e}^\dagger \hat{a} \hat{e}$-type operator" for short). 
In Ref.~\cite{Ichikawa2013}, when we take the expectation value of these terms, we have factorized
the terms into the expectation value of the excitation operator and that of the photon creation or annihilation operator. 
After this approximation, these terms give finite contribution only when an initial photon state is 
a coherent state, which is an eigenstate of the photon annihilation operator.
In particular, they vanish for the photon vacuum state.
To go one step further, we, in this paper, do not perform the above factorization, and solve the time evolution 
equation of the $\hat{e}^\dagger \hat{a} \hat{e}$-type operators simultaneously with that of the excitation 
operator. 
As we show in a later section, this procedure corresponds to counting the self-energy process
of the electron (the electron emits a photon and then absorbs it again), and, consistently, 
it gives non-zero contribution even when the initial state is the photon vacuum state. 

This paper is organized as follows. 
In Sec.~\ref{sec:qnum}, we derive the time evolution equations of the quantum operators. 
After we describe how we expand field operators and define creation and annihilation operators,
we show time derivative of the creation and annihilation operators from the equations of motion of the quantum fields.
Then, we derive the time evolution equation of the excitation operator and $\hat{e}^\dagger \hat{a} \hat{e}$-type operator.
In Sec.~\ref{sec:dm}, we derive time evolution equations of the density matrix
by taking the expectation value of the time derivative of the excitation operator.
We explain our approximation methods to derive closed sets of time evolution equations.
In Sec.~\ref{sec:results}, we show the results of numerical computation of these equations for a hydrogen atom and molecule.
Finally, Sec.~\ref{sec:conclusion} is devoted to our conclusion. 
The notations and conventions follow those in Refs.~\cite{Tachibana2001,Tachibana2003,Tachibana2010,Ichikawa2013}.

%%%%%%%%%%%%%%%%%%%%%%%%%%%%%%%%%%%%%%%
%
\section{Evolution equations for quantum operators}  \label{sec:qnum}
%
%%%%%%%%%%%%%%%%%%%%%%%%%%%%%%%%%%%%%%%

We would like to describe atoms and molecules as systems which consist of electrons, atomic nuclei and photons.
In this paper, we work in the Born-Oppenheimer (BO) approximation, in which positions of the atomic nuclei are fixed.
Then, only electrons and photons are described as quantum field operators, and
atomic nuclei contribute to the charge density as delta functions.
The source of the photon field can be both electrons and atomic nuclei.
The quantum field operators which appear in this paper are 
the four-component Dirac field operator $\hat{\psi}(x)$ for the electron,  
and the $U(1)$ gauge field $\hat{A}_\mu(x)$ for the photon. 
(Incidentally, the approach to treat the electron as the two-component Schr\"{o}dinger field \cite{Tachibana2013}
is also developed in our group. See Refs.~\cite{Senami2013a,Senami2013b,Senami2014} for details.)
Although there are some overlaps in this section with the contents in Ref.~\cite{Ichikawa2013},
we reproduce them in the reorganized form for the convenience of the readers. 

%%%%%%%%%%%%%%%%%%%%%%%%%%%%%%%%%%%%%%%%%%%%%%%%%%%%%%%%
%\subsection{Equations of motion for field operators} \label{sec:qnum_field}
\subsection{Definitions of creation and annihilation operators} \label{sec:qnum_defac}
%%%%%%%%%%%%%%%%%%%%%%%%%%%%%%%%%%%%%%%%%%%%%%%%%%%%%%%%

Our expansion of the Dirac field operator is
\begin{eqnarray}
\hat{\psi}(ct, \vec{r}) = \sum_{n=1}^{N_D}  \sum_{a=\pm}  \hat{e}_{n^a}(t) \psi_{n^a}(\vec{r}), \label{eq:psi_expand2}
\end{eqnarray}
where $\psi_{n^+}(\vec{r})$ and $\psi_{n^-}(\vec{r})$ are respectively the $n$-th electron and positron solutions 
of the four-component Dirac-Hartree-Fock equation under the existence of external electrostatic field, and
they form an orthonormal basis set as
%\begin{eqnarray}
$\int d^3\vec{r}\, \psi_{n^a}^\dagger(\vec{r})\psi_{m^b}(\vec{r}) = \delta_{nm} \delta_{ab}$.  %\label{eq:psi_orthonormality}
%\end{eqnarray}
In Eq.~\eqref{eq:psi_expand2}, $N_D$ is the number of the electron expansion functions, which is same as one of the positron expansion functions.
Considering the Kramers pair, $N_D$ usually equals to twice the number of basis functions used to solve the Dirac equation. 
We note that $N_D$ has to be infinite to make the expansion function set complete,
but this is not available in numerical calculation. 
In practice, we use a finite set which only spans a certain region of the complete space.
We should interpret thus obtained results as phenomena within the subspace spanned by the finite set of expansion functions. 
We may use any localized wavepackets for the expansion functions, but the above choice 
is convenient because we can obtain such functions easily by the publicly available code like DIRAC \cite{DIRAC12}.
The creation and annihilation operators are defined as the coefficients of the expansion functions and 
carry the time dependence. 
In our notation, $\hat{e}_{n^+} $ is the electron annihilation operator and $\hat{e}_{n^-} $ is the positron creation operator. 
Accordingly, $\hat{e}^\dagger_{n^+} $ is the electron creation operator and $\hat{e}^\dagger_{n^-} $ is the positron annihilation operator. 
In the literature, a creation operator is usually expressed by an operator with a dagger as superscript,
but, note that, in our notation, the positron creation operator does not carry a dagger whereas the positron annihilation operator does. 
We adopt this notation to make compact the expression of the expansion of Eq.~\eqref{eq:psi_expand2} and equations below.
The equal-time anti-commutation relation can be written as 
%\begin{eqnarray}
$\left\{ \hat{e}_{n^a}(t), \hat{e}_{m^b}^\dagger(t) \right\} = \delta_{nm} \delta_{ab}$, %\label{eq:ACR_e}
%\end{eqnarray}
and anti-commutators of other combinations are zero.

For later convenience, we here define the electronic excitation operator, which is formed from 
two creation or annihilation operators as
\begin{eqnarray}
\hat{\cal E}_{n^a m^b} \equiv \hat{e}^\dagger_{n^a} \hat{e}_{m^b}.  \label{eq:op_excitation}
\end{eqnarray}
Then, the electronic charge density operator and current density operators are written by the excitation operator as
%By substituting the expansion \eqref{eq:psi_expand2} into Eqs.~\eqref{eq:rhoe} and \eqref{eq:je}, we obtain
\begin{eqnarray}
\hat{\rho}_e(x) &=&  \sum_{n,m=1}^{N_D} \sum_{a,b=\pm} \rho_{n^a m^b}(\vec{r}) \hat{\cal E}_{n^a m^b}(t), \label{eq:op_rhoe} \\
\hat{j}_e^k(x) &=& \sum_{n,m=1}^{N_D} \sum_{a,b=\pm}  j^k_{n^a m^b}(\vec{r}) \hat{\cal E}_{n^a m^b}(t),  \label{eq:op_je}
 \end{eqnarray}
where we define
\begin{eqnarray}
\rho_{n^a m^b}(\vec{r}) &\equiv& (Z_e e) \psi_{n^a}^\dagger(\vec{r})  \psi_{m^b}(\vec{r}),  \\
j^k_{n^a m^b}(\vec{r}) &\equiv& Z_e\, e\,c \left[ \psi_{n^a}^\dagger(\vec{r}) \gamma^0 \gamma^k \psi_{m^b}(\vec{r})  \right], \label{eq:jk}
\end{eqnarray}
and $Z_e=-1$.
The total charge density operator have contribution from both electrons and atomic nuclei,
$\hat{\rho}(x) = \hat{\rho}_e(x) + \sum_{a=1}^{N_n} \hat{\rho}_a(x)$, where  
$\hat{\rho}_a(x)$ is the atomic nuclear charge density operator,
$a$ denotes the type of atomic nucleus and we assume $N_n$ types of atomic nuclei in the system.
Under the BO approximation, $\hat{\rho}_a(x) = Z_a e\, \delta^{(3)}(\vec{r} - \vec{R}_a)$,
where $Z_a$ is the nucleus $a$'s atomic number, and $\vec{R}_a$
should be understood as the direct sum of the position of each nucleus of type $a$ (see Appendix of Ref.~\cite{Ichikawa2013} for details). 
As for the total charge current density operator, since there is no contribution from the atomic nuclei
under the BO approximation, we have $\hat{\vec{j}}(x)=\hat{\vec{j}}_e(x)$. 
Although we do not compute the energy of the system in this paper,
it may be instructive to show the QED Hamiltonian operator. 
This is shown in Appendix \ref{sec:H}.

As for the photon field operator, we use the integral equation form of the equation of motion.
Namely, we use the fact that the formal solutions of the inhomogeneous 
Maxwell equations %\eqref{eq:Maxwell1} and \eqref{eq:Maxwell2} 
are known from the classical electrodynamics 
\cite{Jackson,Loudon,LLCTF},
and express $\hat{A}_\mu(x)$ by such solutions \cite{Tachibana2003,Ichikawa2013}. 
Similar technique is used to derive the so-called Yang-Feldman equation,
which is originally introduced to discuss the S-matrix of QED in Heisenberg picture \cite{Yang1950}. 
%According to this procedure, $\hat{A}_\mu(x)$ is written by the sum of the solution of free Maxwell equations and the particular solution of the inhomogeneous Maxwell equations.
%We denote the former as $\hat{A}_{\rm rad, \mu}(x)$ and the latter as $\hat{A}_{A, \mu}(x)$, 
%so that $\hat{A}_\mu(x) = \hat{A}_{\rm rad, \mu}(x) + \hat{A}_{A, \mu}(x)$.
We adopt the Coulomb gauge, $\vec{\nabla} \cdot \hat{\vec{A}}(x) = 0$.
Then, the scalar potential is given by
\begin{eqnarray}
\hat{A}_0(ct, \vec{r}) = \int d^3\vec{s}\, \frac{\hat{\rho}(ct, \vec{s})}{|\vec{r}-\vec{s}|}, \label{eq:A0}
\end{eqnarray}
and the vector potential by
\begin{eqnarray}
\hat{\vec{A}}(ct, \vec{r}) = \hat{\vec{A}}_{\rm rad}(ct, \vec{r}) + \hat{\vec{A}}_A(ct, \vec{r}), \label{eq:vecA}
\end{eqnarray}
where the first term is the quantized free radiation field in the Coulomb gauge \cite{Loudon,SakuraiQFT} and the second term is the retarded potential. 
$\hat{\vec{A}}_{\rm rad}(ct, \vec{r})$ is expressed as 
\begin{eqnarray}
\hat{A}^k_{\rm rad}(ct, \vec{r})
 &=& \frac{\sqrt{4\pi \hbar^2 c}}{\sqrt{(2\pi\hbar)^3}} \sum_{\sigma=\pm1} \int \frac{d^3 \vec{p}}{\sqrt{2p^0}}
\bigg[ \hat{a}_{\vec{p}_\sigma} e^k(\vec{p},\sigma) e^{-i c p^0 t/\hbar} e^{i \vec{p}\cdot \vec{r} /\hbar} \nonumber \\
& &+ \hat{a}^\dagger_{\vec{p}_\sigma} e^{* k}(\vec{p},\sigma) e^{i c p^0 t/\hbar} e^{-i \vec{p}\cdot \vec{r} /\hbar} \bigg] ,
\label{eq:Aradvec}
\end{eqnarray}
where $\hat{a}_{\vec{p}_\sigma}$ ($\hat{a}^\dagger_{\vec{p}_\sigma}$) are the annihilation (creation) operator
of the photon with momentum $\vec{p}$ and helicity $\sigma$, and
$\vec{e}$ is the polarization vector. They satisfy the commutation relation 
$[ \hat{a}_{\vec{p}_\sigma}, \hat{a}^\dagger_{\vec{q}_\tau} ] = \delta(\vec{p}-\vec{q}) \delta_{\sigma \tau}$,
and commutators of the other combinations are zero. 
Our convention for $\vec{e}(\vec{p},\sigma)$ is found in Ref.~\cite{Ichikawa2013}.
$\hat{\vec{A}}_A(ct, \vec{r})$ is expressed as 
\begin{eqnarray}
\hat{\vec{A}}_A(ct, \vec{r}) = \frac{1}{c} \int d^3\vec{s}\, \frac{\hat{\vec{j}}_T(c u, \vec{s})}{|\vec{r}-\vec{s}|},  \label{eq:A_A}
\end{eqnarray}
where we define the retarded time
$u = t - |\vec{r}-\vec{s}|/c$, and $\hat{\vec{j}}_T(x)$ is the transversal component 
of the current. 
Writing explicitly, 
\begin{eqnarray}
\hat{j}_T^k(cu,\vec{s}) &=&  \sum_{p,q=1}^{N_D} \sum_{c,d=\pm} \left\{ 
 j^k_{p^c q^d}(\vec{s}) \hat{\cal E}_{p^c q^d}(u)  
+ E^k_{p^c q^d}(\vec{s}) \frac{d \hat{\cal E}_{p^c q^d}}{dt}(u)
\right\},   \label{eq:jT}
\end{eqnarray}
where 
\begin{eqnarray}
E^{k}_{n^a m^b}(\vec{R}) 
%\equiv 
%-\frac{1}{4\pi} \frac{\partial}{\partial R^k}V_{n^a m^b}(\vec{R}) 
= 
-\frac{Z_e e}{4\pi} \int d^3\vec{s}\, \psi_{n^a}^\dagger(\vec{s}) \psi_{m^b}(\vec{s})  \frac{(\vec{s}-\vec{R})^k}{|\vec{s}-\vec{R}|^3}.   \label{eq:ef_int} 
\end{eqnarray}

As we mentioned in Sec.~\ref{sec:introduction}, we do not consider the usual asymptotic states of the field operator,
in which the state becomes non-interacting as $t \rightarrow -\infty$ asymptotically.
We assume that the interaction is absent for $t<0$ and the free part expressed above by $\hat{A}_{\rm rad, \mu}(x)$
is considered to realize at $t = 0$. 
This, in particular, implies the initial condition 
\begin{eqnarray}
\hat{j}^\mu(cu,\vec{s}) = 0, \quad u<0,
\end{eqnarray}
and enables us to rewrite the retarded potential as \cite{Tachibana2013}
\begin{eqnarray}
\hat{\vec{A}}_A(ct, \vec{r}) &=& 
\frac{1}{c^2 \pi} \int_{0}^{t }du  \int_{-\infty}^{\infty}d\alpha \exp \left( i\alpha (t-u)^2 \right)  
\int d^3\vec{s}\, \hat{\vec{j}}_T(c u, \vec{s}) \exp \left( -i\alpha  \frac{(\vec{r}-\vec{s})^2}{c^2} \right). 
\label{eq:A_A_mod}
\end{eqnarray}
To derive this, we use the following formulae for the delta function
$\delta(x^2-a^2) = \left\{ \delta(x-a) + \delta(x+a) \right\}/(2a)$ with $a>0$,
and
$\delta( (t-u)^2 - (\vec{r}-\vec{s})^2/c^2 ) = \frac{1}{2\pi} \int_{-\infty}^{\infty} d\alpha \exp [ i\alpha \{ (t-u)^2 -(\vec{r}-\vec{s})^2/c^2\} ]$.
This form may be convenient for the numerical calculation since the retarded time is eliminated,
but at the cost of increasing the dimension of integration. 

In passing, some comments on our assumption at the initial time ($t=0$) may be in order. 
Following Ref.~\cite{Tachibana2005}, we assume that the parameters and fields which
appear in the equations of motion have been renormalized at $t=0$. 
Specifically, the renormalization has been performed in a standard manner as
$m_e = m_{eB} + \delta m_e$, $e = \sqrt{Z_3} e_B$, 
$\hat{\psi}(x) = \hat{\psi}_B(x)/ \sqrt{Z_2}$ and $\hat{A}^\mu(x) = \hat{A}_B^\mu(x)/ \sqrt{Z_3}$,
where ``B" in the subscripts denotes a bare parameter or a bare field,
and $ \delta m_e$, $Z_2$ and $Z_3$ are the renormalization constants. 
However, it does not mean that we do not need renormalization at $t>0$.
In fact, as is discussed later in this paper, the electron mass is shown to be increased by 
including the self-energy process in our time evolution simulation.

%%%%%%%%%%%%%%%%%%%%%%%%%%%%%%%%%%%%%%%%%
\subsection{Time evolution of annihilation and creation operators}  \label{sec:qnum_evac}
%%%%%%%%%%%%%%%%%%%%%%%%%%%%%%%%%%%%%%%%%
The time derivative of $\hat{e}_{n^a}$ is given by substituting the expansion Eq.~\eqref{eq:psi_expand2}
into the Dirac field equation, multiplying by $\psi^\dagger_{n^a}(\vec{r})$, integrating over $\vec{r}$,
and using the orthonormality condition \cite{Ichikawa2013}. 
This leads to
\begin{eqnarray}
\hspace{-0.6cm}
i\hbar \frac{\partial \hat{e}_{n^a}}{\partial t} &=& 
  \sum_{m=1}^{N_D}  \sum_{b=\pm}  h_{n^a m^b} \hat{e}_{m^b}
  +    \sum_{m, p, q=1}^{N_D} \sum_{b, c, d=\pm}(n^a m^b | p^c q^d) \hat{e}^\dagger_{p^c} \hat{e}_{q^d}  \hat{e}_{m^b} 
  \nonumber \\
&-&\frac{1}{c^2} \sum_{m=1}^{N_D}  \sum_{b=\pm}  \sum_{k=1}^3 \int d^3\vec{r}\, d^3\vec{s}\,  
j^k_{n^a m^b}(\vec{r})  \frac{\hat{j}^k_T(c u, \vec{s})}{|\vec{r}-\vec{s}|}  \hat{e}_{m^b} 
 \nonumber \\
&-& \frac{\sqrt{4\pi \hbar^2}}{\sqrt{c (2\pi\hbar)^3}}
 \sum_{m=1}^{N_D}  \sum_{b=\pm} \sum_{k=1}^3  \sum_{\sigma=\pm1} \int \frac{d^3 \vec{p}}{\sqrt{2p^0}} 
\bigg[ 
F^k _{n^a m^b}(\vec{p}) e^k(\vec{p},\sigma) e^{-i c p^0 t/\hbar}\, \hat{a}_{\vec{p}_\sigma} \hat{e}_{m^b} \nonumber \\
& &+F^k_{n^a m^b}(-\vec{p}) e^{* k}(\vec{p},\sigma) e^{i c p^0 t/\hbar}\,   \hat{a}^\dagger_{\vec{p}_\sigma}   \hat{e}_{m^b}
\bigg],
\label{eq:dedt_BO}
\end{eqnarray}
where the various coefficient matrices are defined as follows. 
In the first term, we define
\begin{eqnarray}
h_{n^a m^b} = T_{n^a m^b} + M_{n^a m^b} +   \sum_{a=1}^{N_n} (Z_a e) V_{n^a m^b}(\vec{R}_a),
\label{eq:inth}
\end{eqnarray}
where $T_{n^a m^b}$ is the electronic kinetic energy integral, 
$M_{n^a m^b}$ is the electronic mass energy integral,
and $V_{n^a m^b}(\vec{R})$ is the nuclear attraction integral, respectively defined as
\begin{eqnarray}
T_{n^a m^b} 
&\equiv&  -i\hbar c \int d^3\vec{r}\,\psi_{n^a}^\dagger(\vec{r}) \gamma^0 \vec{\gamma} \cdot \vec{\nabla}  \psi_{m^b}(\vec{r}), \label{eq:kin_int} \\
M_{n^a m^b} 
&\equiv& m_e c^2 \int d^3\vec{r}\,\psi_{n^a}^\dagger(\vec{r}) \gamma^0  \psi_{m^b}(\vec{r}), \label{eq:mass_int} \\
V_{n^a m^b}(\vec{R})
&\equiv& (Z_e  e) \int d^3\vec{s}\, \frac{\psi_{n^a}^\dagger(\vec{s}) \psi_{m^b}(\vec{s}) }{|\vec{s}-\vec{R}|}.   \label{eq:nuc_int}
\end{eqnarray}
They originate respectively in the kinetic term, mass term, and the nuclear charge contribution to the scalar potential term.
%\rem{Here, $Z_a$ is the nucleus $a$'s atomic number, and $\vec{R}_a$
%should be understood as the direct sum of the position of each nucleus of type $a$ (see Appendix of Ref.~\cite{Ichikawa2013} for details). }

In the second term of Eq.~\eqref{eq:dedt_BO}, we define the electronic repulsion integral 
\begin{eqnarray}
(n^a m^b | p^c q^d) 
\equiv
(Z_e e)^2 \int d^3\vec{r}\, d^3\vec{s}\, \psi_{n^a}^\dagger(\vec{r}) \psi_{m^b}(\vec{r}) 
\frac{1}{|\vec{r}-\vec{s}|} \psi_{p^c}^\dagger(\vec{s}) \psi_{q^d}(\vec{s}).
   \label{eq:twoele_int}  
\end{eqnarray}
This term originates in the electronic charge contribution to the scalar potential term.
The third and forth terms of Eq.~\eqref{eq:dedt_BO} originate in the vector potential term. 
The former comes from the retarded part $\hat{\vec{A}}_A$ \eqref{eq:A_A} 
and the latter from the free radiation part $\hat{\vec{A}}_{\rm rad}$ \eqref{eq:Aradvec}.
Finally, in the forth term of Eq.~\eqref{eq:dedt_BO}, we define
\begin{eqnarray}
F^k_{n^a m^b}(\vec{p}) \equiv \int d^3\vec{r}\,  j^k_{n^a m^b}(\vec{r}) e^{i \vec{p}\cdot \vec{r} /\hbar}.
\end{eqnarray}

%%%%%%%%%%%%%%%%%%%%%%%%%%%%%%%%%%%%%%%%%
\subsection{Time evolution of excitation and $\hat{e}^\dagger \hat{a} \hat{e}$-type operators}  \label{sec:qnum_excitation}
%%%%%%%%%%%%%%%%%%%%%%%%%%%%%%%%%%%%%%%%%
In this paper, we are interested in the time evolution of electronic charge density, and its operator $\hat{\rho}_e(x)$ is
expressed by the excitation operator as Eq.~\eqref{eq:op_rhoe}.
Since the excitation operator carries every time dependence of $\hat{\rho}_e(x)$, 
what we need to know is the time evolution equation of the excitation operator. 

The time derivative of the excitation operator $\hat{\cal E}_{n^a m^b}$ \eqref{eq:op_excitation} can be written as
\begin{eqnarray}
\frac{\partial  \hat{\cal E}_{n^a m^b}}{\partial t}
=
 (\hat{O}_{m^b n^a})^\dagger + \hat{O}_{n^a m^b},  \label{eq:dcalEdt}
\end{eqnarray}
where we define
\begin{eqnarray}
\hat{O}_{n^a m^b}  \equiv \hat{e}^\dagger_{n^a}   \frac{\partial \hat{e}_{m^b}}{\partial t}. \label{eq:def_op_O}
\end{eqnarray}
Using Eq.~\eqref{eq:dedt_BO}, this can be readily obtained as
\begin{eqnarray}
i\hbar \hat{O}_{n^a m^b} 
&=&
  \sum_{r=1}^{N_D}  \sum_{e=\pm}  h_{m^b r^e} \hat{e}^\dagger_{n^a} \hat{e}_{r^e}
  +    \sum_{r, p, q=1}^{N_D} \sum_{e, c, d=\pm}(m^b r^e | p^c q^d) \hat{e}^\dagger_{n^a}   \hat{e}^\dagger_{p^c} \hat{e}_{q^d}    \hat{e}_{r^e} 
  \nonumber \\
&-&\frac{1}{c^2} \sum_{r=1}^{N_D}  \sum_{e=\pm}  \sum_{k=1}^3 \int d^3\vec{r}\, d^3\vec{s}\,  
j^k_{m^b r^e}(\vec{r}) \hat{e}^\dagger_{n^a} \frac{\hat{j}^k_T(c u, \vec{s})}{|\vec{r}-\vec{s}|}  \hat{e}_{r^e} 
 \nonumber \\
&-& \frac{1}{\sqrt{2\pi^2 \hbar c}}
 \sum_{r=1}^{N_D}  \sum_{e=\pm} \sum_{\sigma=\pm1} \int \frac{d^3 \vec{p}}{\sqrt{2p^0}} 
 \left[ {\cal F}_{m^b r^e \vec{p}_\sigma}(t) \hat{e}^\dagger_{n^a}\hat{a}_{\vec{p}_\sigma} \hat{e}_{r^e} 
+{\cal F}^*_{r^e m^b \vec{p}_\sigma}(t) \hat{e}^\dagger_{n^a} \hat{a}^\dagger_{\vec{p}_\sigma}   \hat{e}_{r^e}
\right], 
\nonumber \\
\label{eq:op_O}
\end{eqnarray}
where we have defined
\begin{eqnarray}
{\cal F}_{n^a m^b \vec{p}_\sigma}(t) \equiv  \sum_{k=1}^3 F^k _{n^a m^b}(\vec{p}) e^k(\vec{p},\sigma) e^{-i c p^0 t/\hbar}, 
\end{eqnarray}
to make the expression shorter. 
For later convenience, we call the second term of Eq.~\eqref{eq:op_O} ``four-electron term", the third term ``retarded potential term",
and the forth term ``radiation term".

Next, we consider the time derivative of the $\hat{e}^\dagger \hat{a} \hat{e}$-type operators.
Since the time derivative of $\hat{e}^\dagger_{m^b}\hat{a}^\dagger_{\vec{p}_\sigma} \hat{e}_{n^a}$ is known by
taking the Hermite conjugate of that of $\hat{e}^\dagger_{n^a}\hat{a}_{\vec{p}_\sigma} \hat{e}_{m^b}$,
we only show the latter below. 
This can be expressed as
\begin{eqnarray}
\frac{\partial }{\partial t} \left\{ \hat{e}^\dagger_{n^a}\hat{a}_{\vec{p}_\sigma}\hat{e}_{m^b} \right\}
=
(\hat{Q}_{m^b \vec{p}_\sigma n^a})^\dagger + \hat{P}_{n^a \vec{p}_\sigma m^b}, \label{eq:op_deaedt}
\end{eqnarray}
where we define
\begin{eqnarray}
 \hat{P}_{n^a \vec{p}_\sigma m^b} &\equiv& \hat{e}^\dagger_{n^a} \hat{a}_{\vec{p}_\sigma} \frac{\partial\hat{e}_{m^b}}{\partial t},  \label{eq:def_opP} \\
  \hat{Q}_{n^a \vec{p}_\sigma m^b} &\equiv& \hat{e}^\dagger_{n^a} \hat{a}^\dagger_{\vec{p}_\sigma} \frac{\partial\hat{e}_{m^b}}{\partial t}. \label{eq:def_opQ}
\end{eqnarray}
Using Eq.~\eqref{eq:dedt_BO}, we obtain
\begin{eqnarray}
i\hbar \hat{P}_{n^a \vec{p}_\sigma m^b} 
&=&
  \sum_{r=1}^{N_D}  \sum_{e=\pm}  h_{m^b r^e} \hat{e}^\dagger_{n^a} \hat{a}_{\vec{p}_\sigma} \hat{e}_{r^e}
  +    \sum_{r, p, q=1}^{N_D} \sum_{e, c, d=\pm}(m^b r^e | p^c q^d) \hat{e}^\dagger_{n^a}\hat{a}_{\vec{p}_\sigma}  \hat{e}^\dagger_{p^c} \hat{e}_{q^d}    \hat{e}_{r^e} 
  \nonumber \\
&-&\frac{1}{c^2} \sum_{r=1}^{N_D}  \sum_{e=\pm}  \sum_{k=1}^3 \int d^3\vec{r}\, d^3\vec{s}\,  
j^k_{m^b r^e}(\vec{r}) \hat{e}^\dagger_{n^a}\hat{a}_{\vec{p}_\sigma} \frac{\hat{j}^k_T(c u, \vec{s})}{|\vec{r}-\vec{s}|}  \hat{e}_{r^e} 
 \nonumber \\
&-& \frac{1}{\sqrt{2\pi^2 \hbar c}}
 \sum_{r=1}^{N_D}  \sum_{e=\pm} \sum_{\tau=\pm1} \int \frac{d^3 \vec{q}}{\sqrt{2q^0}} \nonumber \\
&& \times  \left[ {\cal F}_{m^b r^e \vec{q}_\tau}(t)  \hat{e}^\dagger_{n^a} \hat{a}_{\vec{p}_\sigma}  \hat{a}_{\vec{q}_\tau} \hat{e}_{r^e} 
+{\cal F}^*_{r^e m^b \vec{q}_\tau}(t)   \hat{e}^\dagger_{n^a} \hat{a}_{\vec{p}_\sigma}  \hat{a}^\dagger_{\vec{q}_\tau}   \hat{e}_{r^e}
\right],  \label{eq:op_P}
\end{eqnarray}
and 
\begin{eqnarray}
i\hbar \hat{Q}_{n^a \vec{p}_\sigma m^b} 
&=&
  \sum_{r=1}^{N_D}  \sum_{e=\pm}  h_{m^b r^e} \hat{e}^\dagger_{n^a} \hat{a}^\dagger_{\vec{p}_\sigma}  \hat{e}_{r^e}
  +    \sum_{r, p, q=1}^{N_D} \sum_{e, c, d=\pm}(m^b r^e | p^c q^d) \hat{e}^\dagger_{n^a}\hat{a}^\dagger_{\vec{p}_\sigma}    \hat{e}^\dagger_{p^c} \hat{e}_{q^d}    \hat{e}_{r^e} 
  \nonumber \\
&-&\frac{1}{c^2} \sum_{r=1}^{N_D}  \sum_{e=\pm}  \sum_{k=1}^3 \int d^3\vec{r}\, d^3\vec{s}\,  
j^k_{m^b r^e}(\vec{r}) \hat{e}^\dagger_{n^a}\hat{a}^\dagger_{\vec{p}_\sigma}   \frac{\hat{j}^k_T(c u, \vec{s})}{|\vec{r}-\vec{s}|}  \hat{e}_{r^e} 
 \nonumber \\
&-& \frac{1}{\sqrt{2\pi^2 \hbar c}}
 \sum_{r=1}^{N_D}  \sum_{e=\pm} \sum_{\tau=\pm1} \int \frac{d^3 \vec{q}}{\sqrt{2q^0}} \nonumber \\
& & \times \left[ {\cal F}_{m^b r^e \vec{q}_\tau}(t)  \hat{e}^\dagger_{n^a} \hat{a}^\dagger_{\vec{p}_\sigma}   \hat{a}_{\vec{q}_\tau}  \hat{e}_{r^e} 
 +{\cal F}^*_{r^e m^b \vec{q}_\tau}(t) \hat{e}^\dagger_{n^a} \hat{a}^\dagger_{\vec{p}_\sigma}   \hat{a}^\dagger_{\vec{q}_\tau}    \hat{e}_{r^e}
\right].\label{eq:op_Q}
\end{eqnarray}
We note that the last term of $\hat{P}_{n^a \vec{p}_\sigma m^b}$ includes the operator product of the form
$\hat{e}^\dagger_{n^a} \hat{a}_{\vec{p}_\sigma}  \hat{a}^\dagger_{\vec{q}_\tau}   \hat{e}_{r^e}$.
This combination of the operators could express a self-energy process, in which the electron emits a photon and then absorbs it again.
The effect of the self-energy process will be discussed and numerically demonstrated in Sec.~\ref{sec:selfenergy}.

%%%%%%%%%%%%%%%%%%%%%%%%%%%%%%%%%%%%%%%%%
%
\section{Evolution equations for density matrix}  \label{sec:dm}
%
%%%%%%%%%%%%%%%%%%%%%%%%%%%%%%%%%%%%%%%%%

%%%%%%%%%%%%%%%%%%%%%%%%%%%%%%%%%%%%%%%%%
\subsection{Definition of density matrix} \label{sec:def_dm} 
%%%%%%%%%%%%%%%%%%%%%%%%%%%%%%%%%%%%%%%%%
We begin by introducing notations regarding expectation values. 
We denote the expectation value of the excitation operator with respect to the Heisenberg initial ket $| \Phi \rangle$ by 
\begin{eqnarray}
{\cal E}_{n^a m^b} \equiv  \langle \Phi |   \hat{\cal E}_{n^a m^b}   | \Phi \rangle,
\end{eqnarray}
and call this quantity the density matrix. 
Below, we sometimes write just brackets around the operator to denote the expectation value with respect to $| \Phi \rangle$,
namely, $\langle  \cdots \rangle \equiv  \langle \Phi | \cdots | \Phi \rangle $ where $\cdots$ stands for some operators.
Since $(\hat{\cal E}_{n^a m^b})^\dagger = \hat{e}^\dagger_{m^b} \hat{e}_{n^a}  = \hat{\cal E}_{m^b n^a}$,
taking the expectation value yields $({\cal E}_{n^a m^b})^* = {\cal E}_{m^b n^a}$, 
showing that the density matrix is a Hermite matrix. 
Similarly, we define the expectation value of Eq.~\eqref{eq:def_op_O} as
${\cal O}_{n^a m^b} \equiv   \langle \Phi |  \hat{O}_{n^a m^b}  | \Phi \rangle$.
Then, since
 $\langle  (\hat{O}_{n^a m^b})^\dagger \rangle = ({\cal O}_{n^a m^b})^* = {\cal O}^\dagger_{m^b n^a}$,
taking the expectation value of Eq.~\eqref{eq:dcalEdt} leads to 
\begin{eqnarray}
\frac{\partial  {\cal E}_{n^a m^b}}{\partial t} =  {\cal O}^\dagger_{n^a m^b} + {\cal O}_{n^a m^b},   \label{eq:ev_dcalEdt}
\end{eqnarray}
which is the time evolution equation for the density matrix. 

Using the density matrix, the expectation value of the electronic charge density operator, Eq.~\eqref{eq:op_rhoe},
can be written as
\begin{eqnarray}
 \langle \Phi |  \hat{\rho}_e(x)| \Phi \rangle =  \sum_{n,m=1}^{N_D} \sum_{a,b=\pm} \rho_{n^a m^b}(\vec{r}) {\cal E}_{n^a m^b}(t).
 \end{eqnarray}
However, this quantity has a non-zero value when the Heisenberg initial ket 
is the vacuum, $| \Phi \rangle = | 0 \rangle$, at $t=0$.
(Remember that $\hat{e}_{n^-} $ is the positron creation operator.)
As is done in the ordinary QED, this can be remedied by computing the 
expectation value of the normal-ordered product of the operators at $t=0$.
In the case of $\hat{\rho}_e(x)$, this can be accomplished by subtracting the vacuum expectation value
of $\hat{\rho}_e(x)$ at $t=0$.
Then, we shall define the electronic charge density $\rho_e(x)$ as the expectation value
of the electronic charge density operator after this subtraction. 
Writing explicitly,
\begin{eqnarray}
\rho_e(x) = \sum_{n,m=1}^{N_D} \sum_{a,b=\pm} \rho_{n^a m^b}(\vec{r}) 
\left\{ {\cal E}_{n^a m^b}(t) - {\cal E}^0_{n^a m^b}(t=0) \right\},   \label{eq:phys_rhoe}
\end{eqnarray}
where we define ${\cal E}^0_{n^a m^b} \equiv  \langle 0|   \hat{\cal E}_{n^a m^b}   | 0 \rangle$.

%%%%%%%%%%%%%%%%%%%%%%%%%%%%%%%%%%%%%%%%%
\subsection{Four-electron term and retarded potential term} \label{sec:approx_4e_ret} 
%%%%%%%%%%%%%%%%%%%%%%%%%%%%%%%%%%%%%%%%%
The four-electron term consists of four electron creation or annihilation operators. 
We approximate this term by 
\begin{eqnarray}
\langle \hat{e}^\dagger_{n^a}  \hat{e}^\dagger_{p^c}    \hat{e}_{q^d}  \hat{e}_{r^e}   \rangle 
&\approx&
\langle \hat{e}^\dagger_{n^a}  \hat{e}_{r^e} \rangle  \langle \hat{e}^\dagger_{p^c}    \hat{e}_{q^d}    \rangle
- \langle \hat{e}^\dagger_{n^a}  \hat{e}_{q^d}  \rangle  \langle \hat{e}^\dagger_{p^c}     \hat{e}_{r^e}   \rangle \nonumber \\
&=&
{\cal E}_{n^a r^e} {\cal E}_{p^c q^d} -{\cal E}_{n^a q^d} {\cal E}_{p^c r^e}. 
\label{eq:approx_4e}
\end{eqnarray}
This decomposition holds exactly for $t=0$ and the approximation is motivated by this fact. 
In our previous paper Ref.~\cite{Ichikawa2013}, we have only used the first term, and
the second term, which describes the exchange effect, has been omitted.

As for the retarded potential term,
since $\hat{j}^k_T$ is expressed as Eq.~\eqref{eq:jT}, it seems to 
have the same structure as the four-electron term.
However, since $\hat{j}^k_T$ is computed at the retarded time and contains a time derivative of 
the excitation operator, the approximation like Eq.~\eqref{eq:approx_4e} is not applicable. 
Thus, we approximate this term by just replacing $\hat{j}^k_T$ by its expectation value as
\begin{eqnarray}
\langle \hat{e}^\dagger_{n^a}(t) \hat{j}^k_T(c u, \vec{s}) \hat{e}_{r^e}(t) \rangle 
\approx
\langle  \hat{j}^k_T(c u, \vec{s})\rangle \langle   \hat{e}^\dagger_{n^a}(t)\hat{e}_{r^e}(t) \rangle 
=
{\cal J}^k_T(c u, \vec{s}) {\cal E}_{n^a r^e}(t),
\label{eq:approx_jT}
\end{eqnarray}
where we define ${\cal J}^k_T \equiv   \langle \Phi | \hat{j}^k_T  | \Phi \rangle$.
With this approximation, 
the two space integrations over $\vec{r}$ and $\vec{s}$ in the expectation value of the retarded term 
becomes that of $c$-number.
In this six-dimensional integration, the retarded time $u$ depends both on $\vec{r}$ and $\vec{s}$,
so the integration has to be done numerically, which is not practical.
To avoid the six-dimensional numerical integration, we rewrite the expression 
into the form in which the retarded time is eliminated.
This is done by the same process as the one used to derive Eq.~\eqref{eq:A_A_mod}.
Then, for the expectation value of the retarded term, we have
\begin{eqnarray}
-\frac{1}{c^2} \sum_{r=1}^{N_D}  \sum_{e=\pm}  \sum_{k=1}^3 \int d^3\vec{r}\, d^3\vec{s}\,  
j^k_{m^b r^e}(\vec{r}) \frac{{\cal J}^k_T(c u, \vec{s})}{|\vec{r}-\vec{s}|}   {\cal E}_{n^a r^e}(t)
= \sum_{r=1}^{N_D}  \sum_{e=\pm}  I_{j_T}[{\cal E},\dot{\cal E}]_{m^b r^e}(t)  {\cal E}_{n^a r^e}(t), \label{eq:approx_O3rd}
\end{eqnarray}
where we define
\begin{eqnarray}
I_{j_T}[{\cal E},\dot{\cal E}]_{m^b r^e}(t) &\equiv&
-\frac{1}{c^3\pi} \sum_{p,q=1}^{N_D}  \sum_{c,d=\pm} \int_{0}^{t }du  \int_{-\infty}^{\infty}d\alpha \exp \left( i\alpha (t-u)^2 \right)  \nonumber \\
& &\times \left\{ 
I_{jj, m^b r^e p^c q^d}(\alpha) {\cal E}_{p^c q^d}(u)  + I_{jE, m^b r^e p^c q^d}(\alpha)  \frac{d {\cal E}_{p^c q^d}}{dt}(u)
\right\},  \label{eq:def_IjT}
\end{eqnarray}
with
\begin{eqnarray}
I_{jj, m^b r^e p^c q^d}(\alpha) &\equiv& 
\sum_{k=1}^3  \int d^3\vec{r}\,  d^3\vec{s}\,  j^k_{m^b r^e}(\vec{r})  j^k_{p^c q^d}(\vec{s}) \exp \left( -i\alpha  \frac{(\vec{r}-\vec{s})^2}{c^2} \right),  
\label{eq:def_Ijj} \\
I_{jE, m^b r^e p^c q^d}(\alpha) &\equiv& 
\sum_{k=1}^3  \int d^3\vec{r}\,  d^3\vec{s}\,  j^k_{m^b r^e}(\vec{r}) E^k_{p^c q^d}(\vec{s}) \exp \left( -i\alpha  \frac{(\vec{r}-\vec{s})^2}{c^2} \right). 
\label{eq:def_IjE}
\end{eqnarray}
Here, we put $[{\cal E},\dot{\cal E}]$ after $I_{j_T}$ for the notation of the integral \eqref{eq:def_IjT}
in order to emphasize that it depends on the density matrix and its time derivative at times earlier than $t$.
Since the functions appearing in Eqs.~\eqref{eq:def_Ijj} and \eqref{eq:def_IjE} are defined by Eqs.~\eqref{eq:jk} and \eqref{eq:ef_int},
$I_{jj}$ and $I_{jE}$ are four-center integrals and analytic formulae can be obtained when the expansion functions
of the Dirac field operator are gaussian functions. 
We show the integral formulae in the Appendix~\ref{sec:molint}.

%%%%%%%%%%%%%%%%%%%%%%%%%%%%%%%%%%%%%%%%%
\subsection{Radiation term} \label{sec:approx_rad} 
%%%%%%%%%%%%%%%%%%%%%%%%%%%%%%%%%%%%%%%%%
In this section, we describe two approximation methods for the expectation value 
of the radiation term. 
We first describe the simplest approximation method, which is same as the one adopted in Ref.~\cite{Ichikawa2013}.
In this approximation, we use
\begin{eqnarray}
\langle \hat{e}^\dagger_{n^a}\hat{a}_{\vec{p}_\sigma}  \hat{e}_{r^e} \rangle 
&\approx&
 \langle \hat{a}_{\vec{p}_\sigma}  \rangle  \langle \hat{e}^\dagger_{n^a} \hat{e}_{r^e} \rangle
= \langle \hat{a}_{\vec{p}_\sigma}  \rangle  {\cal E}_{n^a r^e},
\end{eqnarray}
and, similarly, $\langle \hat{e}^\dagger_{n^a}\hat{a}^\dagger_{\vec{p}_\sigma} \hat{e}_{r^e} \rangle 
\approx \langle \hat{a}_{\vec{p}_\sigma}  \rangle ^* {\cal E}_{n^a r^e}$.
This factorization holds exactly for $t=0$, when $ \hat{e}_{n^a}$ and $\hat{a}_{\vec{p}_\sigma} $ commutes,
and the approximation is motivated by this fact. 
We note that this term may give finite contribution only when the initial photon state is a coherent state,
which is an eigenstate of the photon annihilation operator.
Then, combined with the approximations described in the previous subsection, we obtain
\begin{eqnarray}
i\hbar {\cal O}_{n^a m^b} 
&=&
  \sum_{r=1}^{N_D}  \sum_{e=\pm}  h_{m^b r^e} {\cal E}_{n^a r^e}
  +    \sum_{r, p, q=1}^{N_D} \sum_{e, c, d=\pm}(m^b r^e | p^c q^d) \left( {\cal E}_{n^a r^e} {\cal E}_{p^c q^d} -{\cal E}_{n^a q^d} {\cal E}_{p^c r^e} \right)
  \nonumber \\
&+& \sum_{r=1}^{N_D}  \sum_{e=\pm}  I_{j_T}[{\cal E},\dot{\cal E}]_{m^b r^e}(t)  {\cal E}_{n^a r^e}
 \nonumber \\
&-& \frac{1}{\sqrt{2\pi^2 \hbar c}}
 \sum_{r=1}^{N_D}  \sum_{e=\pm} \sum_{\sigma=\pm1} \int \frac{d^3 \vec{p}}{\sqrt{2p^0}} 
\left[ {\cal F}_{m^b r^e \vec{p}_\sigma}(t) \langle \hat{a}_{\vec{p}_\sigma} \rangle  {\cal E}_{n^a r^e} 
+{\cal F}^*_{r^e m^b \vec{p}_\sigma}(t)\langle \hat{a}_{\vec{p}_\sigma} \rangle ^* {\cal E}_{n^a r^e}
\right]. \label{eq:calO_1} \nonumber \\
\end{eqnarray}
This gives us a closed differential equation for the density matrix. 

In the second approximation method, we do not use the above factorization.
We use evolution equation for the expectation value of the $\hat{e}^\dagger \hat{a} \hat{e}$-type operators
simultaneously with one for the density matrix. 
We first define
\begin{eqnarray}
{\cal E}_{n^a \vec{p}_\sigma m^b} \equiv  \langle \Phi |   \hat{e}^\dagger_{n^a} \hat{a}_{\vec{p}_\sigma}  \hat{e}_{m^b}    | \Phi \rangle.
\end{eqnarray}
Note that $\langle \Phi |   \hat{e}^\dagger_{n^a} \hat{a}^\dagger_{\vec{p}_\sigma} \hat{e}_{m^b} | \Phi \rangle = {\cal E}^*_{m^b \vec{p}_\sigma n^a}$.
We next define the expectation value of the operator $\hat{P}_{n^a \vec{p}_\sigma m^b}$, Eq.~\eqref{eq:def_opP},
as ${\cal P}_{n^a \vec{p}_\sigma m^b} \equiv   \langle \Phi |  \hat{P}_{n^a \vec{p}_\sigma m^b}  | \Phi \rangle$.
Similarly for the operator $\hat{Q}_{n^a \vec{p}_\sigma m^b}$, Eq.~\eqref{eq:def_opQ}, we define ${\cal Q}_{n^a \vec{p}_\sigma m^b} \equiv   \langle \Phi |  \hat{Q}_{n^a \vec{p}_\sigma m^b}  | \Phi \rangle$.
Then, the expectation value of Eq.~\eqref{eq:op_O} can be written as
\begin{eqnarray}
i\hbar {\cal O}_{n^a m^b} 
&=&
  \sum_{r=1}^{N_D}  \sum_{e=\pm}  h_{m^b r^e} {\cal E}_{n^a r^e}
  +    \sum_{r, p, q=1}^{N_D} \sum_{e, c, d=\pm}(m^b r^e | p^c q^d) \left( {\cal E}_{n^a r^e} {\cal E}_{p^c q^d} -{\cal E}_{n^a q^d} {\cal E}_{p^c r^e} \right)
  \nonumber \\
&+& \sum_{r=1}^{N_D}  \sum_{e=\pm}  I_{j_T}[{\cal E},\dot{\cal E}]_{m^b r^e}(t)  {\cal E}_{n^a r^e}
 \nonumber \\
&-& \frac{1}{\sqrt{2\pi^2 \hbar c}}
 \sum_{r=1}^{N_D}  \sum_{e=\pm}  \sum_{\sigma=\pm1} \int \frac{d^3 \vec{p}}{\sqrt{2p^0}} 
\left[ {\cal F}_{m^b r^e \vec{p}_\sigma}(t) {\cal E}_{n^a \vec{p}_\sigma r^e}  
+{\cal F}^*_{r^e m^b \vec{p}_\sigma}(t)  {\cal E}^*_{r^e \vec{p}_\sigma n^a}
\right], \label{eq:calO_2} \nonumber \\
\end{eqnarray}
and the time evolution of  ${\cal E}_{n^a \vec{p}_\sigma m^b}$ can be expressed as
\begin{eqnarray}
\frac{\partial {\cal E}_{n^a \vec{p}_\sigma m^b}}{\partial t}
= 
{\cal Q}^*_{m^b \vec{p}_\sigma n^a} + {\cal P}_{n^a \vec{p}_\sigma m^b}, \label{eq:ev_deaedt}
\end{eqnarray}
which is obtained by taking the expectation value of Eq.~\eqref{eq:op_deaedt}.

As for ${\cal P}_{n^a \vec{p}_\sigma m^b}$, the expectation value of the operator product in the second term of Eq.~\eqref{eq:op_P} is approximated to be
\begin{eqnarray}
\langle \hat{e}^\dagger_{n^a} \hat{a}_{\vec{p}_\sigma}  \hat{e}^\dagger_{p^c} \hat{e}_{q^d}    \hat{e}_{r^e}  \rangle
\approx
\langle  \hat{a}_{\vec{p}_\sigma} \rangle \left\{ 
\langle \hat{e}^\dagger_{n^a}  \hat{e}_{r^e} \rangle  \langle \hat{e}^\dagger_{p^c}    \hat{e}_{q^d}    \rangle
- \langle \hat{e}^\dagger_{n^a}  \hat{e}_{q^d}  \rangle  \langle \hat{e}^\dagger_{p^c}     \hat{e}_{r^e}   \rangle 
\right\}. 
\end{eqnarray}
In this approximation, we assume that the initial photon state is a number state and not a coherent state.
Then, since $\langle  \hat{a}_{\vec{p}_\sigma} \rangle=0$, the contribution of this term vanishes. 
The third term of Eq.~\eqref{eq:op_P} includes the retarded time and its expectation value would be
approximated in the same way as the retarded potential term of Eq.~\eqref{eq:op_O}.
Following the procedure described below Eq.~\eqref{eq:approx_jT},
we have the expression of the form of Eq.~\eqref{eq:approx_O3rd} with ${\cal E}_{n^a r^e}(t)$
replaced by ${\cal E}_{n^a \vec{p}_\sigma r^e}(t)$.
The forth term of Eq.~\eqref{eq:op_P} includes two four-operator terms, which are approximated to be
\begin{eqnarray}
\langle \hat{e}^\dagger_{n^a} \hat{a}_{\vec{p}_\sigma}  \hat{a}_{\vec{q}_\tau} \hat{e}_{r^e} \rangle
 &\approx& 
\langle  \hat{a}_{\vec{p}_\sigma}  \hat{a}_{\vec{q}_\tau}  \rangle \langle \hat{e}^\dagger_{n^a} \hat{e}_{r^e} \rangle,  
 \\
\langle \hat{e}^\dagger_{n^a} \hat{a}_{\vec{p}_\sigma}  \hat{a}^\dagger_{\vec{q}_\tau}   \hat{e}_{r^e} \rangle
 &\approx&
 \langle  \hat{a}_{\vec{p}_\sigma}  \hat{a}^\dagger_{\vec{q}_\tau}  \rangle  \langle \hat{e}^\dagger_{n^a} \hat{e}_{r^e} \rangle.
\end{eqnarray}
Since we do not consider a coherent state for the initial photon state as mentioned above, 
$\langle  \hat{a}_{\vec{p}_\sigma}  \hat{a}_{\vec{q}_\tau}  \rangle=0$  in the first equation
and its contribution vanishes. 
In the second equation, since 
$\langle  \hat{a}_{\vec{p}_\sigma}  \hat{a}^\dagger_{\vec{q}_\tau}  \rangle
= (n_{\vec{p}_\sigma}+1) \delta^{(3)}(\vec{p}-\vec{q}) \delta_{\sigma \tau}$,
where $n_{\vec{p}_\sigma}$ is the occupation number of the photon mode $(\vec{p},\sigma)$ 
in the initial state, it may give a non-zero contribution.
Putting these approximations together, 
we have
\begin{eqnarray}
i\hbar {\cal P}_{n^a \vec{p}_\sigma m^b} &=&
  \sum_{r=1}^{N_D}  \sum_{e=\pm}  h_{m^b r^e} {\cal E}_{n^a  \vec{p}_\sigma r^e}
  + \sum_{r=1}^{N_D}  \sum_{e=\pm}  I_{j_T}[{\cal E},\dot{\cal E}]_{m^b r^e}(t)  {\cal E}_{n^a \vec{p}_\sigma  r^e}
 \nonumber \\
 &-& \frac{1}{\sqrt{2\pi^2 \hbar c}}
 \sum_{r=1}^{N_D}  \sum_{e=\pm} \frac{1}{\sqrt{2p^0}} {\cal F}^*_{r^e m^b \vec{p}_\sigma}(t) (n_{\vec{p}_\sigma}+1) {\cal E}_{n^a  r^e},
 \label{eq:calP}
\end{eqnarray}
when we assume a number state for the photon initial state.
The expectation value of Eq.~\eqref{eq:op_Q} can be approximated in a similar manner to be
\begin{eqnarray}
i\hbar {\cal Q}_{n^a \vec{p}_\sigma m^b} &=&
  \sum_{r=1}^{N_D}  \sum_{e=\pm}  h_{m^b r^e} {\cal E}^*_{r^e  \vec{p}_\sigma n^a}
  + \sum_{r=1}^{N_D}  \sum_{e=\pm}  I_{j_T}[{\cal E},\dot{\cal E}]_{m^b r^e}(t) {\cal E}^*_{r^e  \vec{p}_\sigma n^a}
 \nonumber \\
 &-& \frac{1}{\sqrt{2\pi^2 \hbar c}}
 \sum_{r=1}^{N_D}  \sum_{e=\pm} \frac{1}{\sqrt{2p^0}} {\cal F}_{m^b r^e \vec{p}_\sigma}(t) n_{\vec{p}_\sigma} {\cal E}_{n^a  r^e},
 \label{eq:calQ}
\end{eqnarray}
where we have used 
$\langle  \hat{a}^\dagger_{\vec{p}_\sigma}  \hat{a}_{\vec{q}_\tau}  \rangle
= n_{\vec{p}_\sigma} \delta^{(3)}(\vec{p}-\vec{q}) \delta_{\sigma \tau}$.

%%%%%%%%%%%%%%%%%%%%%%%%%%%%%%%%%%%%%%%%%
%
\section{Results}  \label{sec:results}
%
%%%%%%%%%%%%%%%%%%%%%%%%%%%%%%%%%%%%%%%%%
In this section, we show the results of numerical solution of the time evolution equations 
which have been derived in the previous section.
The computation is performed for a hydrogen atom and molecule using 
the QEDynamics code \cite{QEDynamics} developed in our group.
In Sec.~\ref{sec:calc_setup}, we describe our setups for numerical calculation
including the initial condition for the density matrix.
The results of two approximation methods discussed in Sec.~\ref{sec:approx_rad}
are respectively presented in Sec.~\ref{sec:coherent} and Sec.~\ref{sec:selfenergy}. 

%%%%%%%%%%%%%%%%%%%%%%%%%%%%%%%%%%%%%%%%%
\subsection{Setups for numerical calculation} \label{sec:calc_setup} 
%%%%%%%%%%%%%%%%%%%%%%%%%%%%%%%%%%%%%%%%%
To perform  numerical calculation, we first need to determine an orthonormal set of 
expansion functions to define the electron creation and annihilation operators,
as explained in Sec.~\ref{sec:qnum_defac}.
We generate the set by solving the Dirac equation with the four-component Dirac-Coulomb hamiltonian.
They are computed by the publicly available program package DIRAC \cite{DIRAC12},
using the Hartree-Fock method with the STO-3G basis set.
Note that, for the hydrogen atom in this basis set, we have two ($=N_D$) orbitals for
electron and positron respectively taking into account the Kramers partners.
The density matrix is 4 $\times$ 4 matrix whose components denote
electron ($1^+$), its Kramers partner ($\bar{1}^+$), positron ($1^-$), and its Kramers partner ($\bar{1}^-$).
As shown here, we put a bar on the orbital number to denote the Kramers partner.
Similarly, the size of the density matrix for the hydrogen molecule in this basis set is 8 ($N_D=4$).

We next explain the initial condition for the density matrix. 
We choose the initial Heisenberg ket $|\Phi \rangle$ to be the ground states of the hydrogen atom and molecule,
which are obtained by the above explained computation method.
Namely, expressing $|\Phi \rangle = |\Phi_e \rangle \otimes  |\Phi_{ph} \rangle$,
where $|\Phi_e \rangle$ is the electron part and $|\Phi_{ph} \rangle$ is the photon part, 
we use $|\Phi_e \rangle = \hat{e}^\dagger_{1^+}|0\rangle$ for the hydrogen atom,
and $|\Phi_e \rangle = \hat{e}^\dagger_{\bar{1}^+} \hat{e}^\dagger_{1^+}|0\rangle$ for the hydrogen molecule.
In general, the ground state of a $N_e$-electron system in the Hartree-Fock method is expressed as
$|\Phi_e \rangle = \prod_{i=1}^{N_e/2} \hat{e}^\dagger_{\bar{i}^+} \hat{e}^\dagger_{i^+}  |0\rangle$
when $N_e$ is even, and 
$|\Phi_e \rangle = \hat{e}^\dagger_{((N_e+1)/2)^+} \prod_{i=1}^{(N_e-1)/2} \hat{e}^\dagger_{\bar{i}^+} \hat{e}^\dagger_{i^+}|0\rangle$ when $N_e$ is odd.
For later use, we here introduce the terminology ``occupied" orbitals. 
If $|\Phi_e \rangle$ contains $\hat{e}^\dagger_{i^+}$, the $i$-th electron orbital is called ``occupied"
($i$ can be with or without bar),
and we can write $|\Phi_e \rangle = \prod_{i={\rm occupied}} \hat{e}^\dagger_{i^+}  |0\rangle$.
Then, using the anti-commutation relation, %Eq.~\eqref{eq:ACR_e}, 
the initial condition for the density matrix is 
\begin{eqnarray}
{\cal E}_{n^a m^b}(t=0) &=& \left\{ \begin{array}{ll}
\delta_{nm} & (a=b=+,\, n:{\rm occupied}) \\
\delta_{nm} & (a=b=-) \\
0 & ({\rm otherwise}) \\
\end{array} \right. .
\end{eqnarray}
In particular, the vacuum expectation value at $t=0$, which is needed to compute Eq.~\eqref{eq:phys_rhoe}, is 
\begin{eqnarray}
{\cal E}^0_{n^a m^b}(t=0)  &=& \left\{ \begin{array}{ll}
\delta_{nm} & (a=b=-) \\
0 & ({\rm otherwise}) \\
\end{array} \right. .
\end{eqnarray}

Other numerical details are as follows. 
We work in the atomic units so that $m_e = e = \hbar = 1$, and $c = 137.035 999 679$.
The 1\,a.u. of time corresponds to 2.419$\times 10^{-17}$\,s or 24.19\,as.
As for the positions of the atomic nuclei, we locate them at the origin in the case of the hydrogen atom,
and at $(x, y, z)=(0,0,\pm0.7)$ in the case of the hydrogen molecule. 
For both hydrogen atom and molecule, we report the electronic charge density at $(x, y, z)=(0,0,1)$.
To solve the differential equations, we use the Euler method with the time step $10^{-9}$\,a.u. 
In this paper, we omit the contribution from the retarded potential
by setting the integral $I_{j_T}[{\cal E},\dot{\cal E}]$, Eq.~\eqref{eq:def_IjT}, to be zero. 
This integral, including numerical integration, has to be computed
at every time step, and performing this straightforwardly takes too much computational time. 
We shall study an effective approximation method in our future work, and just neglect it in the present work.

%%%%%%%%%%%%%%%%%%%%%%%%%%%%%%%%%%%%%%%%%
\subsection{Effect of photon coherent state} \label{sec:coherent} 
%%%%%%%%%%%%%%%%%%%%%%%%%%%%%%%%%%%%%%%%%
In this section, we show the results when the first approximation method explained 
in Sec.~\ref{sec:approx_rad} is adopted.
Namely, we solve the time differential equation \eqref{eq:ev_dcalEdt} using Eq.~\eqref{eq:calO_1}.
As mentioned in the end of Sec.~\ref{sec:calc_setup}, the third term of Eq.~\eqref{eq:calO_1},
expressing the retarded potential, is neglected. 
As for the photon initial Heisenberg ket, when there is no radiation field, $|\Phi_{ph} \rangle = | 0 \rangle $,
since $\langle \hat{a}_{\vec{p}_\sigma} \rangle = 0$, the forth term of Eq.~\eqref{eq:calO_1} is dropped. 
In fact, $\langle \hat{a}_{\vec{p}_\sigma} \rangle = 0$ holds when $|\Phi_{ph} \rangle$ is any photon number state.
We may have $\langle \hat{a}_{\vec{p}_\sigma} \rangle \neq 0$ when $|\Phi_{ph} \rangle$ is a coherent state,
and we study its effect in this section. 
Since we quantize the radiation field in the whole space, we consider the continuous-mode coherent state.
Following the notation of Ref.~\cite{Loudon}, we denote it as $| \{ \alpha \} \rangle$.
This is characterized by the eigenvalues of the photon annihilation operator of each mode $(\vec{p},\sigma)$
as
\begin{eqnarray}
\hat{a}_{\vec{p}_\sigma} | \{ \alpha \} \rangle = \alpha(\vec{p},\sigma) | \{ \alpha \} \rangle,
\end{eqnarray}
where $\alpha(\vec{p},\sigma)$ is called spectral amplitude \cite{Loudon}. 
In this paper, we use the delta-function type spectral amplitude and its center is chosen 
to be a mode $(\vec{p}_j, \sigma_j)$, as
\begin{eqnarray}
 \alpha(\vec{p},\sigma) = \alpha_j \delta^{(3)}(\vec{p} - \vec{p}_j) \delta_{\sigma \sigma_j}. \label{eq:sa_deltafunc}
\end{eqnarray}
Such photon state corresponds to a classical oscillating electromagnetic field whose
propagating direction is $\vec{p}_j$, direction of circularly polarization is $\sigma_j$, and 
amplitude is proportional to $\alpha_j$.
The period of the oscillations is determined by $p^0 = |\vec{p}_j|  $ as $2\pi/(p^0 c)$.

\begin{figure}
\begin{center}
\includegraphics[width=12cm]{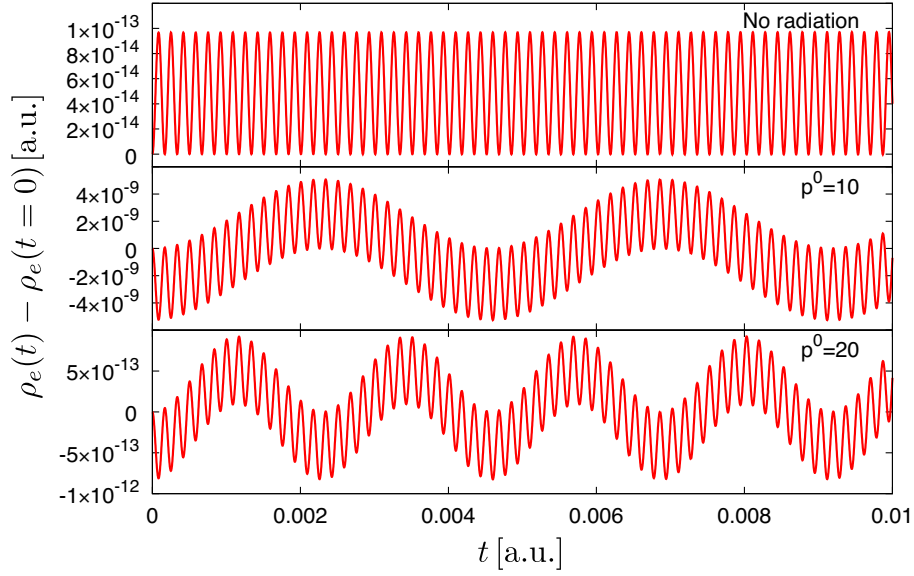}
\caption{Time evolution of the charge density of the hydrogen atom at $(x, y, z) = (0, 0, 1)$. 
The variation from the initial value is plotted. 
The upper panel shows the result when there is no photon in the initial state.
In the middle and lower panels, the initial photon states are chosen to be coherent states
with the photon modes whose energy is $p^0 = 10$ and 20 respectively. 
In both cases, they are chosen to be circularly polarized in the positive direction and have 
momenta in the direction of $x$-axis positive. See the texts for other details. 
}
\label{fig:H_coh}
\end{center}
\end{figure}

\begin{figure}
\begin{center}
\includegraphics[width=12cm]{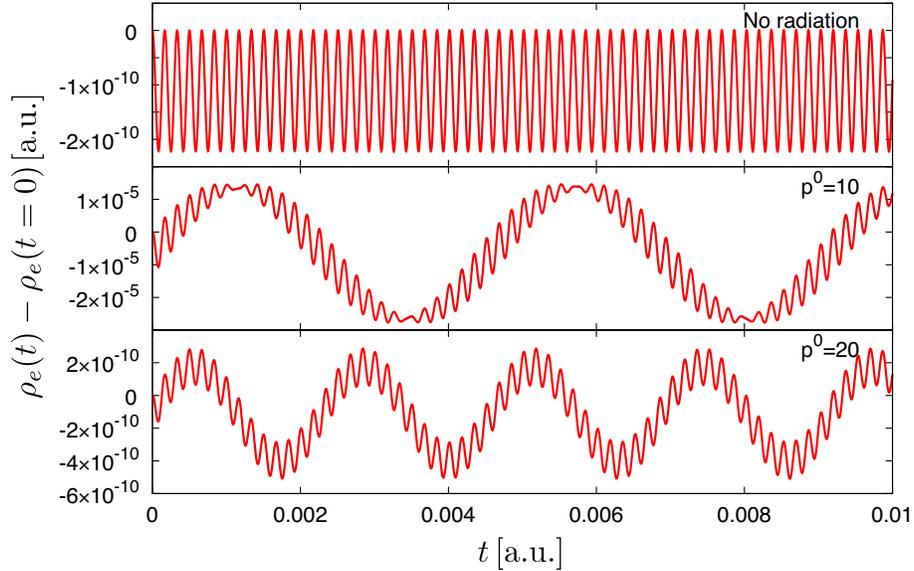}
\caption{Similar to Fig.~\ref{fig:H_coh} for the hydrogen molecule. 
}
\label{fig:H2_coh}
\end{center}
\end{figure}

We first show the results when there is no initial radiation field.
The case of the hydrogen atom is shown in the upper panel of Fig.~\ref{fig:H_coh}
and the case of the hydrogen molecule is in the upper panel of Fig.~\ref{fig:H2_coh}.
In these figures, the variation of the electronic charge density from its initial value is plotted. 
The common feature is the oscillations with very short period of about $1.7 \times 10^{-4}$\,a.u. 
As has been argued in Ref.~\cite{Ichikawa2013}, since this is very close to the period which
is determined from twice the mass of electron, $2\pi/(2m_e c^2)=1.67 \times 10^{-4}$,
it can be interpreted as the fluctuations originated from virtual electron-positron pair 
creations and annihilations. 
Hence, we call this phenomenon the ``electron-positron oscillations" \cite{Ichikawa2013}.
In Ref.~\cite{Ichikawa2013}, we show this using the hydrogen atom. 
In this paper, we show that the electron-positron oscillations occur similarly for the hydrogen
molecule, and expect that we find them universally for any atomic and molecular systems.  

We next show the results when the initial photon state is a coherent state.
We choose its spectral amplitude to be the form expressed by Eq.~\eqref{eq:sa_deltafunc}
with $\vec{p}_j/ |\vec{p}_j| =  (1,0,0)$ and $\sigma_j = +1$.
We compute the cases with $p^0 =  10$ and 20 for each hydrogen atom and molecule. 
The case of $p^0=10$ (20) is shown in the middle (lower) panel in Figs.~\ref{fig:H_coh} and \ref{fig:H2_coh}.
In these panels, we can see that the electron-positron oscillations with the short period are
modulated by the longer period oscillations which are caused by the external oscillating 
electromagnetic field.
In fact, the periods of the modulating oscillations seen in the panels for $p^0=10$ in both hydrogen atom and molecule
are close to $2\pi/(p^0 c)=4.59 \times 10^{-3}$\,a.u.
Similarly, in the panels for $p^0=20$, we see the periods of the modulating oscillations are close to  
$2\pi/(p^0 c)=2.29 \times 10^{-3}$\,a.u.
We note that we have tuned the value of $\alpha_j$ for each case, in order to make
these effects visible clearly. 
We have chosen $\alpha_j = 10^3$ ($10^4$) when $p^0=10$ (20) for the hydrogen atom
and $\alpha_j = 2 \times 10^4$ ($10^5$) when $p^0=10$ (20) for the hydrogen molecule. 

We note that whether this ``electron-positron oscillations" is a real physical phenomenon 
or an artifact of our model is an open question.
We can say with certainty that the electron-positron oscillations are caused by including the positron solutions
in our expansion functions for the Dirac field operator
(so they do not take place if we model the electron by the Schr\"{o}dinger field),
and, in fact, $(1^+,1^-)$-component of the density matrix oscillates in the case of the hydrogen atom simulation. 
However, it is also true that the results are obtained with several approximations and 
some ingredients of QED are missing. 
In particular, including the retarded potential may affect the electron-positron oscillations.

%%%%%%%%%%%%%%%%%%%%%%%%%%%%%%%%%%%%%%%%%
\subsection{Effect of electron self-energy} \label{sec:selfenergy} 
%%%%%%%%%%%%%%%%%%%%%%%%%%%%%%%%%%%%%%%%%
In this section, we show the results when the second approximation method explained 
in Sec.~\ref{sec:approx_rad} is adopted. 
Namely, we solve time differential equations \eqref{eq:ev_dcalEdt} and \eqref{eq:ev_deaedt}
using Eqs.~\eqref{eq:calO_2}, \eqref{eq:calP} and \eqref{eq:calQ}.
As in Sec.~\ref{sec:coherent}, the terms which originate from the retarded potential are neglected.
As for the photon initial Heisenberg ket, we assume that there is no radiation field, $|\Phi_{ph} \rangle = | 0 \rangle $,
in this section. 
Therefore, the occupation number is zero for every photon mode $(\vec{p},\sigma)$, $n_{\vec{p}_\sigma}=0$.
However, even in such a case, as is expressed by the factor $(n_{\vec{p}_\sigma}+1)$ in the third term of Eq.~\eqref{eq:calP},
every photon mode contributes to the radiation term.
This is reasonable because,
as is mentioned in the end of Sec.~\ref{sec:qnum_excitation}, this term comes from 
a self-energy process of the electron, in which the electron emits a virtual photon and then absorbs it again.
This virtual photon could have any momentum.

Before showing our results, we explain here the numerical details regarding the discretization of the photon modes.
We have to discretize the index $\vec{p}_\sigma$ in ${\cal E}_{n^a \vec{p}_\sigma m^b}$ to perform
numerical calculation.
We adopt the spherical coordinate system $(p^0, \theta, \phi)$ to express $\vec{p}$ and
use equally spaced grid points for each coordinate whose numbers are denoted by $N_{p^0}$, $N_\theta$ and $N_\phi$ respectively.
One more parameter we need to specify is the maximum of $p^0$, denoted by $p^0_{\rm max}$.
We first set $N_\theta=5$ and $N_\phi=4$, and compare the case with $(p^0_{\rm max}, N_{p^0}) = (10, 10)$ and 
the case with $(p^0_{\rm max}, N_{p^0}) = (20, 20)$.
Since the results do not change, we adopt $(p^0_{\rm max}, N_{p^0}) = (10, 10)$ in the following. 
As for the choice of $(N_\theta, N_\phi)$, the result with $(N_\theta, N_\phi) =(7, 8)$ is slightly different from
the case with $(N_\theta, N_\phi) =(5, 4)$, but it is almost same as the result with $(N_\theta, N_\phi) =(11, 10)$.
Therefore, in summary, we adopt $(p^0_{\rm max}, N_{p^0}, N_\theta, N_\phi) =(10, 10, 11, 10)$
as the photon-mode discretization parameters. 

\begin{figure}
\begin{center}
\includegraphics[width=12cm]{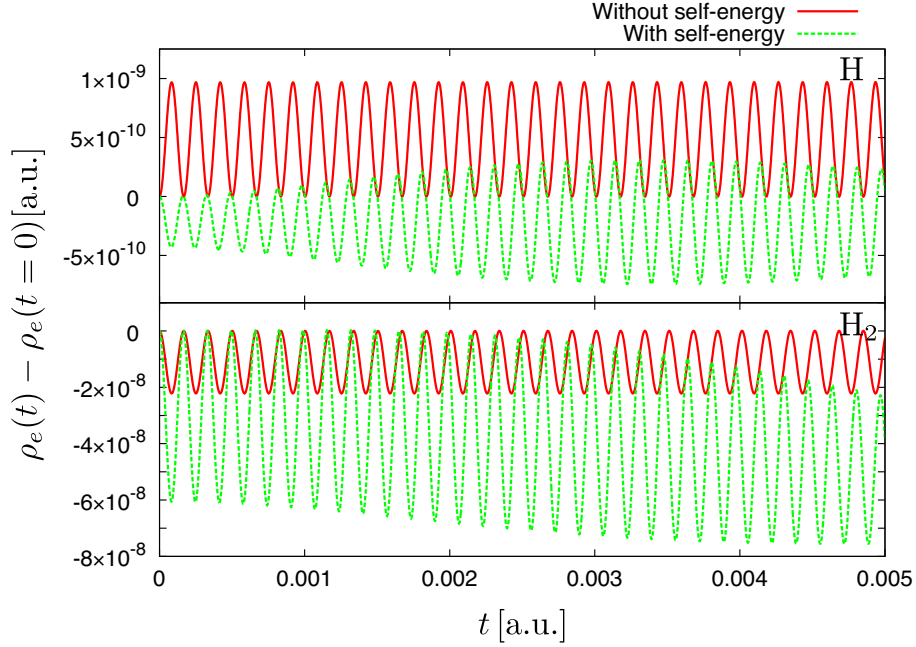}
\caption{Time evolution of the charge density of the hydrogen atom (upper panel) and molecule (lower panel) at $(x, y, z) = (0, 0, 1)$. 
The variation from the initial value is plotted. There is no photon in the initial state.
The cases without and with the self-energy process are respectively plotted by the red solid lines and green dashed lines 
in each panel. 
As for the case without the self-energy process, to make comparison easily, we plot the values
which are multiplied by $10^4$ for H and by $10^2$ for H$_2$.
}
\label{fig:se}
\end{center}
\end{figure}

The results are shown in Fig.~\ref{fig:se}, the upper panel for the hydrogen atom and the lower panel for the hydrogen molecule. 
As in Sec.~\ref{sec:coherent}, the variation of the electronic charge density from its initial value is plotted. 
In each panel, the solid red line shows the result without the self-energy process and the green dashed line shows 
one including the self-energy process. 
Note that computation for the case without the self-energy process is same as the one described in Sec.~\ref{sec:coherent}.
However, in Fig.~\ref{fig:se}, the results are multiplied by $10^4$ for the hydrogen atom and by $10^2$ for the hydrogen molecule
in order to make easy the comparison with the case including the self-energy process. 
In the figure, we see the electron-positron oscillations still take place when we include the self-energy process
for both hydrogen atom and molecule. 
However, the period of the oscillations is slightly shorter than the case without the self-energy process. 
We have mentioned earlier that this rapid oscillations are originated from virtual electron-positron pair 
creations and annihilations and their period is inversely proportional to the electron mass. 
Therefore, decrease in the period implies increase in the electron mass. 
This is reasonable because the electron mass should be increased by including the self-energy process. 

We note that this ``increase" in the electron mass is not the physical reality. 
The self-energy of the electron is the interaction energy between the electron and the electromagnetic field
 which is originated from the electron itself \cite{Jackson,SakuraiQFT}.
(It exists for either classical or quantum electrodynamics.)
Since this is something we cannot remove from the electron, the total energy 
including the self-energy is considered to give the observed electron mass. This is the idea of the (mass) renormalization.
Although the method of the renormalization is well-established for the ordinary perturbative QED, 
since it is based on the notion of the asymptotic states, which exist in the infinite past and future,
it is not straightforwardly applicable to our QED simulation in which finite time evolution is followed. 
We have succeeded in extracting the self-energy of the electron in our simulation.
Our next task is how to renormalize the ``increase" in the electron mass, and this will be studied in our future work.

In the end of this section, we shall make comments on the self-energy effect.
Although the change in the period of the electron-positron oscillations can be interpreted as 
the electron mass shift due to the self-energy effect, 
one may wonder why the shift is not infinite as in the ordinary QED. 
One reason is that we have truncated the infinitely many hierarchy of time evolution equations 
of operators at the level of $\hat{e}^\dagger \hat{a} \hat{e}$-type operator. 
We would have a greater self-energy effect by considering the time evolution equation of
higher order operators such as the $\hat{e}^\dagger  \hat{a} \hat{a}^\dagger \hat{e}$-type operator,
which appears in the last term of Eq.~\eqref{eq:op_P}.
Another reason is that we have only included localized wavepackets for the expansion functions in Eq.~\eqref{eq:psi_expand2}.
During the self-energy process, when the electron emits the virtual photon, 
the electron could be in an unbounded state as a virtual particle, but such a state cannot be expressed by
our present expansion functions. 
In order to improve this point, we may add plane wave functions to the expansion functions to express continuum modes.

%%%%%%%%%%%%%%%%%%%%%%%%%%%%%%%%%%%%%%%%%
%
\section{Conclusion}  \label{sec:conclusion}
%
%%%%%%%%%%%%%%%%%%%%%%%%%%%%%%%%%%%%%%%%%

In this paper, we have discussed a method to follow the step-by-step time evolution of atomic and molecular systems based on 
QED. 
Our strategy includes expanding the electron field operator by localized wavepackets
to define creation and annihilation operators
and following the time evolution using the equations of motion of the field operator in the Heisenberg picture.
Under the BO approximation, 
we have first derived a time evolution equation for the excitation operator, which is the product of two creation or annihilation operators.
We need this operator to construct operators of physical quantities such as the electronic charge density operator. 
We have then described our approximation methods to obtain time differential equations of the electronic density matrix,
which is defined as the expectation value of the excitation operator.

In particular, we have presented two approximation methods for the expectation value 
of the radiation term, which includes the $\hat{e}^\dagger \hat{a} \hat{e}$-type operators.
One is to factorize their expectation values into the expectation value of the excitation operator and that of the photon creation
or annihilation operator,
and has been used to study the effect of external oscillating electromagnetic field by setting
the initial photon state as a coherent state.
Another is to solve the time evolution 
equation of the $\hat{e}^\dagger \hat{a} \hat{e}$-type operators simultaneously with that of the excitation 
operator, which enables us to include the self-energy effect of the electron. 
By solving these equations numerically,
we have shown the electron-positron oscillations appear in the charge density of a hydrogen atom and molecule,
for the cases both with and without including the self-energy process. 
We have also shown that the period of the electron-positron oscillations becomes shorter by including the self-energy process,
and it can be interpreted as the increase in the electron mass due to the self-energy. 

Although the results obtained in this paper can be reasonably interpreted so far, 
there are many things to incorporate for establishing the time evolution simulation method of 
atomic and molecular systems based on QED.
Two important points which are not included in the present work are computation of the retarded potential 
and renormalization of the electron mass.
As for the retarded potential, its computation is likely to be achieved by using the gaussian integral formulae
which have been derived in the Appendix, 
although we need efficient approximation and storage methods.
As for the electron mass renormalization, we shall develop a different method from that of the ordinary QED,
because our renormalization should also be performed step-by-step in time. 
Specifically, we may need a time-dependent renormalization factor. 
These issues will be addressed in our future works and incorporated in our computation code.

\noindent 
%%%%%%%%%%%%%%%%%%%%%%%%%%%%%%%%%%%%
\section*{Acknowledgment}
%%%%%%%%%%%%%%%%%%%%%%%%%%%%%%%%%%%%
%Theoretical calculations were partly performed using Research Center for
% Computational Science, Okazaki, Japan.
This work is supported by Grant-in-Aid for Scientific research (No.~25410012)
from the Ministry of Education, Culture, Sports, Science and Technology, Japan.

%%%%%%%%%%%%%%%%%REFERENCES%%%%%%%%%%%%%%%%%%%%%%%%%

%%%%%%%%%%%%%%%%%REFERENCES%%%%%%%%%%%%%%%%%%%%%%%%%

\appendix

%%%%%%%%%%%%%%%%%%%%%%%%%%%%%%%%%%%%%%%%%
%
\section{QED Hamiltonian operator and energy}  \label{sec:H}
%
%%%%%%%%%%%%%%%%%%%%%%%%%%%%%%%%%%%%%%%%%
The QED Hamiltonian density operator $\hat{H}_{\rm QED}(\vec{r})$ \cite{Tachibana2003,Tachibana2010} can be expressed by a sum of the electromagnetic field energy density operator $\hat{H}_\gamma(\vec{r})$ 
and the energy density operator of electron $\hat{H}_e(\vec{r})$ (Ref.~\cite{Tachibana2010}, Eqs.~(3.4) and (3.5)).
The QED Hamiltonian operator can be written as 
\begin{eqnarray}
\int d^3\vec{r} \hat{H}_{\rm QED}(\vec{r}) 
&=&
%\int d^3\vec{r}\, \Bigg[ \frac{1}{2}  \hat{A}_0 \hat{\rho} +  \frac{1}{8\pi} \left(\frac{1}{c} \frac{\partial \hat{\vec{A}}}{\partial t}  \right)^2 + \frac{1}{8\pi} \left( \vec{\nabla} \times \hat{\vec{A}} \right)^2 \nonumber \\
\int d^3\vec{r}\, \Bigg[ \frac{1}{2}  \hat{A}_0 \hat{\rho} +  \frac{1}{8\pi} \left(\frac{1}{c} \frac{\partial \hat{\vec{A}}}{\partial t}  \right)^2 - \frac{1}{8\pi}  \hat{\vec{A}} \cdot \nabla^2  \hat{\vec{A}}  \nonumber \\
%& & + c \hat{\bar{\psi}} \left\{ -i\hbar \gamma^k \left( \partial_k + i\frac{Z_e e}{\hbar c} \hat{A}_k \right) + m_e c \right\} \hat{\psi} \Bigg],\nonumber \\
& & + c \hat{\bar{\psi}} \left\{ -i\hbar \vec{\gamma} \cdot \left( \vec{\nabla} - i\frac{Z_e e}{\hbar c} \hat{\vec{A}} \right) + m_e c \right\} \hat{\psi} \Bigg],
\end{eqnarray}
in the Coulomb gauge. 
This can be expressed using the creation and annihilation operators by substituting Eqs.~\eqref{eq:psi_expand2}, \eqref{eq:A0}, \eqref{eq:vecA}, \eqref{eq:Aradvec}, and \eqref{eq:A_A}.
Since the terms involving $\hat{\vec{A}}$ cannot be put in a simpler form due to the existence of the retarded time, 
we show the Hamiltonian operator expressed by the creation and annihilation operators under the electrostatic limit, namely $\hat{\vec{A}}=0$,
\begin{eqnarray}
& &\int d^3\vec{r} \hat{H}_{\rm QED, electrostatic}(\vec{r}) \nonumber \\
&=& 
\sum_{n,m=1}^{N_D} \sum_{a,b=\pm}  h_{n^a m^b} \hat{\cal E}_{n^a m^b}  
+\frac{1}{2} \sum_{n,m,p,q=1}^{N_D} \sum_{a,b,c,d=\pm}(n^a q^d | m^b p^c) \hat{e}^\dagger_{n^a} \hat{e}^\dagger_{m^b} \hat{e}_{p^c} \hat{e}_{q^d},
\label{eq:Hes}
\end{eqnarray}
where the coefficient matrices are defined in Eqs.~\eqref{eq:inth} and \eqref{eq:twoele_int}.
Here, we have excluded some terms which are infinite constants. 

By taking the expectation value of the normal-ordered product of Eq.~\eqref{eq:Hes}
with respect to the Heisenberg ket, 
we can obtain the energy of the system under the electrostatic limit, $E_{\rm QED, electrostatic}$.
When the Heisenberg ket is assumed to be the one introduced in Sec.~\ref{sec:calc_setup}, at $t=0$,
it gives the ordinary DHF energy as
\begin{eqnarray}
E_{\rm QED, electrostatic} 
=
\sum_{n=1}^{N_D} \sum_{a=\oplus}  h_{n^a n^a} + \frac{1}{2}  \sum_{n,m=1}^{N_D} \sum_{a,b=\oplus}  \left\{  (n^a n^a  | m^b m^b) - (n^a m^b | m^b n^a) \right\},
\end{eqnarray}
where $\oplus$ denotes the occupied electron orbitals. 

%%%%%%%%%%%%%%%%%%%%%%%%%%%%%%%%%%%%%%%%%
%
\section{Molecular integral formulae for retarded potential term}  \label{sec:molint}
%
%%%%%%%%%%%%%%%%%%%%%%%%%%%%%%%%%%%%%%%%%
When we compute the retarded potential term as described in Sec.~\ref{sec:approx_4e_ret},
we need two types of four-center integrals shown in Eqs.~\eqref{eq:def_Ijj} and \eqref{eq:def_IjE}.
To our knowledge, gaussian integral formulae which are needed to compute them are not seen in literature. 
In the present paper, the retarded potential term is neglected as explained in Sec.~\ref{sec:results},
but we shall derive the formulae in this section for convenience of future works.
The formulae and derivation here are based on a method described in Ref.~\cite{McMurchie1978}.

Let us write an unnormalized gaussian function whose center is on $\vec{A}$ and exponent is $\alpha_A$ as
\begin{eqnarray}
%\tilde{g}(\vec{r}; \vec{A},\alpha_A,\vec{n}) &\equiv& (x-A_x)^{n_x} (y-A_y)^{n_y} (z-A_z)^{n_z} e^{-\alpha_A r^2_A}, 
\tilde{g}(\vec{r}; \vec{A},\alpha_A,\vec{n}) &\equiv& (x-A_x)^{n_x} (y-A_y)^{n_y} (z-A_z)^{n_z} e^{-\alpha_A |\vec{r} - \vec{A} |^2}.
\end{eqnarray}
%where $r^2_A = (x-A_x)^2 +  (y-A_y)^2 + (z-A_z)^2$.
Then, we need the integral of the form
\begin{eqnarray}
\int d^3\vec{r}\,  d^3\vec{s}\,  \tilde{g}(\vec{r}; \vec{R}_i, \alpha_i, \vec{n}_i) \tilde{g}(\vec{r}; \vec{R}_j, \alpha_j, \vec{n}_j) 
 \tilde{g}(\vec{s}; \vec{R}_k, \alpha_k, \vec{n}_k) \tilde{g}(\vec{s}; \vec{R}_l, \alpha_l, \vec{n}_l) \theta(\vec{r}, \vec{s}),   \label{eq:gausstwoele}
\end{eqnarray}
with $\theta(\vec{r}, \vec{s})$ being
\begin{eqnarray}
\theta_{jj}(\vec{r}, \vec{s}; \alpha) \equiv \exp \left( -i\alpha  \frac{|\vec{r}-\vec{s}|^2}{c^2} \right),  \label{eq:thetajj}
\end{eqnarray}
for Eq.~\eqref{eq:def_Ijj}, and 
\begin{eqnarray}
\theta^k_{jE}(\vec{r}, \vec{s}; \alpha) \equiv \int d^3\vec{t}\,   \left\{ \frac{\partial}{\partial t^k}\frac{1}{|\vec{s}-\vec{t}|} \right\} \exp \left( -i\alpha  \frac{|\vec{r}-\vec{t}|^2}{c^2} \right), \label{eq:thetajE}
\end{eqnarray}
for Eq.~\eqref{eq:def_IjE}.
Note that this becomes the usual electronic repulsion integral when $\theta(\vec{r}, \vec{s}) =  1/|\vec{r}-\vec{s}|$.
In the method of Ref.~\cite{McMurchie1978}, in order to compute the integral of the form Eq.~\eqref{eq:gausstwoele},
we first need to compute
\begin{eqnarray}
[000|\theta|000] = \int d^3\vec{r}\,  d^3\vec{s}\,  \exp\left(-\alpha_P  |\vec{r} - \vec{P} |^2  \right) \exp\left(-\alpha_Q  |\vec{s} - \vec{Q} |^2  \right) \theta(\vec{r}, \vec{s}),
\end{eqnarray}
where 
\begin{eqnarray}
 \alpha_P &=& \alpha_i+\alpha_j, \\
\vec{P} &=& \frac{\alpha_i \vec{R}_i + \alpha_j \vec{R}_j}{\alpha_i+ \alpha_j}, \\
 \alpha_Q &=& \alpha_k+\alpha_l, \\
\vec{Q} &=& \frac{\alpha_k \vec{R}_k + \alpha_l \vec{R}_l}{\alpha_k+ \alpha_l}, 
\end{eqnarray}
and then compute
\begin{eqnarray}
& &\left[ NLM | \theta | N' L' M' \right] \nonumber \\
&& = \left(\frac{\partial}{\partial P_x}\right)^N \left(\frac{\partial}{\partial P_y}\right)^L \left(\frac{\partial}{\partial P_z}\right)^M
\left(\frac{\partial}{\partial Q_x}\right)^{N'} \left(\frac{\partial}{\partial Q_y}\right)^{L'} \left(\frac{\partial}{\partial Q_z}\right)^{M'}\left[ 000 | \theta | 000 \right].
\end{eqnarray}
%This means that if we have a formula for $s$-type gaussian functions, one for higher angular momentum gaussian functions 
%can be obtained by 
Finally, Eq.~\eqref{eq:gausstwoele} is obtained by summation over $N$, $L$, $M$, $N'$, $L'$ and $M'$ after
multiplying  $\left[ NLM | \theta | N' L' M' \right]$ by appropriate coefficients which depend on $\vec{n}_i$, $\vec{n}_j$, $\vec{n}_k$ and $\vec{n}_l$.
Thus, we show below $[000|\theta|000]$ and $\left[ NLM | \theta | N' L' M' \right]$ for 
$\theta_{jj}$ and $\theta^k_{jE}$.
%$\theta_{jj}$ (Eq.~\eqref{eq:thetajj})  and $\theta^k_{jE}$ (Eq.~\eqref{eq:thetajE}).

We first consider the case of $\alpha = 0$. 
As for $\theta_{jj}$, since  $\theta_{jj}(\vec{r}, \vec{s}; \alpha=0) =1$, 
Eq.~\eqref{eq:gausstwoele} is the product of two overlap integrals. 
As for $\theta_{jE}$, since  $\theta_{jE}(\vec{r}, \vec{s}; \alpha=0) =0$, Eq.~\eqref{eq:gausstwoele} is also zero.

In the case of  $\alpha \neq 0$, we can show that
\begin{eqnarray}
\left[ 000 | \theta_{jj} | 000 \right] &=& \pi^3 B^{-3/2} \exp \left( -\alpha_T |\vec{D}|^2 \right), \\
\left[ 000 | \theta^k_{jE} | 000 \right] &=& -4 \pi^4 B^{-3/2} F_1(\alpha_T  |\vec{D}|^2) D^k,
\end{eqnarray}
where 
\begin{eqnarray}
\vec{D} &=& \vec{P} -\vec{Q}, \\
A &=& \frac{i\alpha}{c^2}, \\
B &=&  A(\alpha_p + \alpha_q) +  \alpha_p \alpha_q, \\
C &=& \alpha_p \alpha_q A, \\
\alpha_T &=& \left(  \frac{1}{\alpha_p} + \frac{1}{\alpha_q} + \frac{1}{A} \right)^{-1} = \frac{C}{B},
\end{eqnarray}
and
\begin{eqnarray}
F_j(T) = \int_0^1 u^{2j} \exp\left( -Tu^2 \right) du,
\end{eqnarray}
is a function defined in Ref.~\cite{McMurchie1978} and its recursion formula is also discussed there. 

The differentiation to derive $\left[ NLM | \theta | N' L' M' \right]$ can be done in a straightforward manner. 
As for $\theta_{jj}$, we can show
\begin{eqnarray}
\left[ NLM | \theta_{jj} | N' L' M' \right] &=& \pi^3 B^{-3/2} \exp \left( -\alpha_T |\vec{D}|^2 \right) \alpha_T^{\frac{N+L+M+N'+L'+M'}{2}}(-1)^{N+L+M} \nonumber \\
&\times& H_N(\alpha_T^{1/2} D_x) H_L(\alpha_T^{1/2} D_y) H_M(\alpha_T^{1/2} D_z) \nonumber \\
&\times& H_{N'}(\alpha_T^{1/2} D_x) H_{L'}(\alpha_T^{1/2} D_y) H_{M'}(\alpha_T^{1/2} D_z),
\end{eqnarray}
where $H_n(x)$ is a Hermite polynomial of degree $n$.
As for $\theta_{jE}$, we can show
\begin{eqnarray}
\hspace{-2cm}
\left[ NLM | \theta^x_{jE} | N' L' M' \right] &=& 
  -4 \pi^4 B^{-3/2}  (-1)^{N'+L'+M'}  \nonumber \\ 
  &\times&  \left\{ D_x  \tilde{R}_{N+N', L+L', M+M'} + (N+N') \tilde{R}_{N+N'-1, L+L', M+M'} \right\}, \\
\left[ NLM | \theta^y_{jE} | N' L' M' \right] &=&  
  -4 \pi^4 B^{-3/2}  (-1)^{N'+L'+M'} \nonumber \\
  &\times & \left\{ D_y  \tilde{R}_{N+N', L+L', M+M'} + (L+L') \tilde{R}_{N+N', L+L'-1, M+M'} \right\}, \\
\left[ NLM | \theta^z_{jE} | N' L' M' \right] &=&  
  -4 \pi^4 B^{-3/2}  (-1)^{N'+L'+M'}\nonumber \\  
  &\times & \left\{ D_z  \tilde{R}_{N+N', L+L', M+M'} + (M+M') \tilde{R}_{N+N', L+L', M+M'-1} \right\}.
\end{eqnarray}
Here, we have defined
\begin{eqnarray}
\tilde{R}_{NLM} %&=& \left(\frac{\partial}{\partial a}\right)^N \left(\frac{\partial}{\partial b}\right)^L \left(\frac{\partial}{\partial c}\right)^M \int_0^1 u^2 e^{-Tu^2} du
%=  \left(\frac{\partial}{\partial a}\right)^N \left(\frac{\partial}{\partial b}\right)^L \left(\frac{\partial}{\partial c}\right)^M F_1(T), 
=  \left(\frac{\partial}{\partial D_x}\right)^N \left(\frac{\partial}{\partial D_y}\right)^L \left(\frac{\partial}{\partial D_z}\right)^M F_1(T), 
\end{eqnarray}
where $T = \alpha_T (D_x^2+D_y^2+D_z^2)$.
%In our case, $\alpha = \alpha_T$, $a = D_x$, $b = D_y$, and $c = D_z$.
For generating a table of all $\tilde{R}_{NLM}$ up to some maximum $N+L+M$, recursion relations discussed in Ref.~\cite{McMurchie1978} can be applied. 
In particular, we can use the recursion relation for the more general integral $R_{NLMj}$, 
\begin{eqnarray}
R_{NLMj} &=&
% (-\alpha^{1/2})^{N+L+M}(-2\alpha)^j \nonumber \\
%&& \times \int_0^1 u^{N+L+M+2j} H_N(\alpha^{1/2} a u)H_L(\alpha^{1/2} b u)H_M(\alpha^{1/2} c u) e^{-Tu^2} du,
 (-\alpha_T^{1/2})^{N+L+M}(-2\alpha_T)^j \nonumber \\
&& \times \int_0^1 u^{N+L+M+2j} H_N(\alpha_T^{1/2} D_x u)H_L(\alpha_T^{1/2} D_y u)H_M(\alpha_T^{1/2} D_z u) e^{-Tu^2} du,
\end{eqnarray}
%which has been defined above Eq.~(4.4) of Ref.~\cite{McMurchie1978},
through the relation
\begin{eqnarray}
\tilde{R}_{NLM} = \frac{R_{NLM 1}}{-2\alpha_T}.
\end{eqnarray}
The details of the recursion relations and efficient numerical techniques are found in Ref.~\cite{McMurchie1978}.

%%%%%%%%%%%%%%%%%%%%%%%%%%%%%%%%%%%%%%%%%%%%%%%%%%%%%%%%%%%%%%%

\newpage

\end{document}